\documentclass[%
 reprint,onecolumn,
 notitlepage,showkeys,
superscriptaddress,nofootinbib,
 amsmath,amssymb,
 aps,
prd,
]{revtex4-2}
\usepackage{epsfig}
\usepackage{hyperref}
\usepackage{orcidlink}
\hypersetup{           
    colorlinks=true,                
    breaklinks=true,                
    urlcolor= red,                
    linkcolor= black,                
    bookmarksopen=false,
    filecolor=black,
    citecolor=blue,
    linkbordercolor=blue}
\usepackage{graphicx}
\usepackage{orcidlink}
\usepackage{multirow}
\usepackage{times}
\usepackage[utf8]{inputenc}

\begin{document}
\title{Stability of the Einstein Static Universe in Zero-Point Length Cosmology with Topological Defects}

\author{Kalyan Bhuyan \orcidlink{0000-0002-8896-7691}}%
 \email{kalyanbhuyan@dibru.ac.in}
\affiliation{%
 Department of Physics, Dibrugarh University, Dibrugarh \\
 Assam, India, 786004}%
\affiliation{Theoretical Physics Divison, Centre for Atmospheric Studies, Dibrugarh University, Dibrugarh, Assam, India 786004}

\author{Mrinnoy M. Gohain \orcidlink{0000-0002-1097-2124}}
\email{mrinmoygohain19@gmail.com}
\thanks{(Corresponding Author)}
\affiliation{%
 Department of Physics, Dibrugarh University, Dibrugarh \\
 Assam, India, 786004}


\begin{abstract}
Recently, zero-point length cosmology has shown some positive insights into some non-singular aspects of the early Universe. In addition, topological defects are known to play a significant role by its presence as a part of the total energy in the very early Universe. We investigate the stability issue of the Einstein static phase in the emergent scenario of the Universe in a generalized framework of zero-point length cosmology in the presence of topological defects in the very early times. We derive the modified Friedmann equations, where the matter sector includes an extra energy density term arising from $n$-dimensional topological defects. We have studied the possibility of graceful exit of emergent scenario and its stability using dynamical system analysis and against homogeneous scalar perturbation. We also analysed the stability against inhomogeneous density perturbation, vector perturbation and tensor perturbation. Through the stability analysis, it has been shown that the model parameters associated with zero-point length setting and $n$-dimensional topological defects play a visible role in the phase transition process from the ESU to the inflationary regime. Also, interestingly it is found that there exists a mutual interplay between the zero-point length parameter, and the dimension of topological defect on the stability of the ESU on the basis of inhomogeneous density perturbation. Finally, the stability is also tested against vector and tensor perturbation, which shows that the ESU is stable against such perturbations.
\end{abstract}

\maketitle
\tableofcontents
\section{Introduction}
\label{intro}
According to general relativity (GR), matter and spacetime were concentrated at the beginning of the universe in a pointlike singularity with infinite density, sometimes called the Planck's regime or the Big-bang singularity or the initial singularity.  The laws of classical physics disintegrate at this point, making a theory of quantum gravity essential.  The shortcomings of classical GR made it necessary to develop alternative theories—or, more specifically, theories of quantum gravity based on the gravitational sector's principles of quantum mechanics—in order to better understand the characteristics of the initial singularity. In order to solve the initial singularity problem, quantum gravity (and hence quantum cosmology) has become quite popular.  Examples of potential solutions to the initial singularity problem include ekpyrotic/cyclic, bouncing universe, and string theory-based cosmological theories.  Furthermore, Ellis et al. \cite{ellis1, ellis2} proposed the so-called ``\emph{emergent Universe}" (EU) scenario in GR with a closed Friedmann-Lemaitre-Robertson-Walker (FLRW) universe as a novel solution to the initial singularity problem in the early Universe.  A phase transition into the standard inflationary cosmology occurs after the Universe resides in an ever-existing Einstein static (ES) phase. The EU is an improvement over the standard inflationary Universe, which is basically singularity-free. As a result, the ESU takes the place of the initial singularity issue in an EU.  Ellis et al. \cite{ellis1,ellis2} proposed the existence of the ESU in the infinite past as a different solution to the initial singularity problem in their first publication.  The feasibility of a closed FLRW geometry with positive spatial curvature is necessary for this formulation.  As is common in classic inflationary cosmology, the model describes a transition from the ES phase to a relatively brief period of inflation, followed by a reheating phase. 

To model this EU scenario, Ellis and his colleagues used a scalar field, a feasible potential, and a flat potential.  The so-called "horizon problem" is automatically resolved by the original static universe without the need for additional physical mechanisms.  The model must have a stable ESU and a seamless transition to the inflationary stage (graceful exit) in order to avoid the initial singularity.  In order for the scenario to effectively avoid a singularity, both of these requirements must be met.  The EU is unable to resolve the singularity issue if one of the requirements is not met. The former condition was not met by Ellis et al.'s model, which posed a serious stability issue.  Since, Barrow et al. \cite{Barrow2003May} found that the ESU in GR is unstable under perturbations, their original version of the EU scenario constructed in GR failed to avoid the initial singularity issue. 

However, in the early universe, physically motivated modifications, for example, semi-classical or quantum corrections, entropy corrections, dimensional corrections etc. may tilt the scales in support of the emergent scenario. To put it in simple terms, although the emergent scenario collapses in the GR framework, string inspired corrections, based on T-duality and also known as zero-point length correction may be a key factor in the mitigation of the initial singularity problem. This correction could potentially ameliorate the circumstance of the issue. This idea has inspired some crucial works on the non-singular cosmological sector, with an intent to attain promising results as opposed to GR. Very recently, Sheykhi et al \cite{Sheykhi2024Jul} studied the possibility of a non-singular Universe through modifying the gravitational potential by including zero-point length correction and the Verlinde's theory of entropic force. They proposed that the consideration of zero-point length correction to the GR field equations can naturally  alleviate the initial singularity of the early Universe. 

In this context, it so appears that the stability problem can be resolved if one deals with corrected models of GR and entropic force cosmology. Following entropy force cosmology, the stability issue of the ESU is studied by Heydarzade et al \cite{Heydarzade2017Jan}, in the background of deformed Horava-Lifshitz cosmology. There are various methods, including perturbation approaches like homogeneous and inhomogeneous perturbations and dynamical systems analysis, can be employed to study the stability of the ESU. For instance, Huang et al \cite{Huang2015Jan} showed that the ESU is unstable in \( f(G) \) Gauss-Bonnet gravity, in both open and closed spatial geometry where inhomogeneous perturbations destabilizes the solution. Khodadi et al \cite{Khodadi2022Jun} studied the EU scenario in Energy-Momentum Squared Gravity, where they showed that  high-energy corrections can lead to a stable ESU. Atazadeh and Darabi \cite{Atazadeh2017Jun} examined the ESU under GUP effects, where they demonstrated that the presence of a minimal length arising from GUP allows for dynamic stability and stability under various perturbations. Heydarzade and Darabi \cite{Heydarzade2015Apr} analyzed the stability of the ESU in the framework of induced matter brane gravity, where they showed that a positive spatial curvature permits stability under scalar and tensor perturbations, whereas vector perturbations show neutral stability. 
There also have been several works on EU related to the primordial power spectrum, for example  Labra\~{n}a \cite{Labrana2015Apr} studied superinflation in a EU, where he showed that it suppresses the large-scale CMB anisotropies, that explains the observed absence of power at the large angular scales. Huang et al \cite{Huang2022Dec,Huang2023Aug} studied the CMB signature of the EU, and showed that the ESU succeeded by an ultraslow-roll inflationary phase gives rise to damped CMB TT-spectrum at large scales, which generates spectra that are very much similar to the ones produced by ultraslow-roll inflation. Martineau and Barrau \cite{Martineau2018Dec} analyzed the primordial tensor power spectrum in EU through a toy model, where they investigated the conditions under which a scale-invariant spectrum can be obtained. Palermo et al \cite{Palermo2022Dec} studied the primordial tensor power spectrum within EU framework, including Continuous Spontaneous Localization (CSL) to correct tensor perturbations and their evolution into inflation.

The Emergent scenario is an alternative framework to address the issue of the initial singularity, which is as a non-singular version of the usual inflationary Universe, succeeding a self sustaining ESU. To stay out of the quantum gravity domain, the ESU's initial size must be larger than Planck's scale.  Notable aspects of the ESU include the absence of the problems concerning quantum gravity and the lack of an initial singularity. For the EU scenario to be implemented successfully, the ESU's stability is necessary.  Several frameworks including alternative theories of gravity like massive gravity\cite{Parisi2012Jul,Li2019May,Paul2018Feb,Mousavi2017Jun}, mimetic gravity \cite{Huang2020Sep}, Eddington inspired Born Infeld gravity \cite{Li2017Jul}, scalar tensor theories of gravity, Rastall theories\cite{Darabi2018Jul,Shabani2022Sep}, Brans Dicke theories\cite{delCampo2007Nov,Labrana2019Apr,Labrana2021Aug,delCampo2009Jul}, scalar-tensor theories \cite{Miao2016Oct}, Lyra geometry \cite{Darabi2015Aug}, Gauss-Bonnet gravity\cite{Huang2015Jan,Gohain2024Jun}, Einstein-Cartan theories \cite{Huang2015May,Shabani2019Mar,Hadi2018Jan}, higher-dimensional theories \cite{Paul2020Aug}, braneworld theories\cite{Zhang2016Jul,Heydarzade2016Jun,Heydarzade2015Apr}, inclusion of anisotropy effects \cite{Ghorani2021Jun}, $f(R,T)$ gravity \cite{Sharif2019Dec,Gohain2023Mar}, $f(Q)$ theories \cite{Gohain2024Feb}, theories with natural UV cutoffs \cite{Khodadi2017Dec}, Ho\v{r}ava-Lifshitz $f(R)$ theory \cite{Khodadi2016Jun}, doubly-general relativity \cite{Khodadi2015Dec},scalar fluid cosmology \cite{Khodadi2018Jun,Bohmer2015Dec}, energy-momentum squared gravity \cite{Khodadi2022Jun}, theories with GUP correction \cite{Atazadeh2017Jun}, non-minimal derivative coupling \cite{Huang2018Jan}, gravitational particle creation \cite{Gangopadhyay2016Oct} etc have been used to study thoroughly the stability in relation to the EU.

Building on these motivations, the paper is based on the implementation of a successful EU in a combined background of zero-point length cosmology and topological defects, which is arranged as follows: In section \ref{review}, we provide a review the basic framework of zero-point length cosmology and the importance of topological defects in cosmology, using which we formulate the necessary Friedmann equations. In section \ref{esu_eqs} using dynamical systems, we shall study the stability aspects and the graceful exit mechanism in our model. In section\ref{hom_scalar_pert}, we study the stability aspects of the ESU with respect to homogeneous scalar perturbations. In section \ref{inhomo_dens}, we discuss the stability in terms of inhomogeneous density perturbations. In section \ref{vec_tens_pert}, we analyse the stability with respect to vector and tensor perturbation and finally in section \ref{conc} we summarize the results of the study.

\section{Brief Review of Zero-point Length Cosmology}
\label{review}
In recent days, the laws of thermodynamics are quite extensively studied in cosmological systems as they can be analogously interpreted with the equations of gravitational systems. This possibility has been explored with rigour over the last few decades, like in the independent works of Jacobson and Padmanabhan  \cite{Jacobson1995Aug,Padmanabhan2005Jan,
Padmanabhan2010Mar}. They developed the mathematical foundation of gravity-thermodynamics correspondence which laid the ground for further exploration of this interesting possibility. The interaction of gravity and the thermodynamical rules has been studied extensively in many settings. In the theoretical framework of FLRW cosmology, the cosmological Friedmann equations which govern the universe's evolution at the apparent horizon are notably comparable to the first law of thermodynamics which can be particularly useful in the context of the early and late Universe. 

Singularity resolution has also been quite extensively investigated in both black hole physics and cosmology. For instance, fundamental theories of quantum gravity like string theory has been employed to find regular black hole solutions and dealing with the resolution of the initial singularity of the early Universe \cite{Sheykhi2024Jul}. The concept of T-duality in string theory leads to a method for addressing singularity resolution, in which a momentum space propagator derived from the path integral duality has been suggested. For a particle of mass $m_0$, the propagator is given by \cite{Padmanabhan1997Mar,Smailagic2003Aug,
Spallucci2005Aug,Fontanini2006Feb}
\begin{equation}
G(p)=-\frac{l_0}{\sqrt{p^2+m_0^2}} K_1\left(l_0 \sqrt{p^2+m_0^2}\right),
\label{prop1}
\end{equation}
where $p^2 = \hbar^2 \kappa^2$ is the momentum squared, $K_1 (x)$ is the modified Bessel function of the second kind and $l_0$ is the zero point length of spacetime. For massless particles, Eq. \eqref{prop1} takes form (setting $\hbar = 1$) \cite{Nicolini2019Oct}
\begin{equation}
G(k) = - \frac{l_0}{\sqrt{\kappa^2}}K_1 \left(l_0 \sqrt{\kappa^2}\right).
\label{prop2}
\end{equation}
The corresponding Newtonian potential at distance $r$ can be found as \cite{Nicolini2019Oct}
\begin{equation}
V(r) = - \frac{M}{\sqrt{r^2 + l_0^2}},
\label{potential}
\end{equation}
where $M$ is the mass of the source generating the potential. According to \cite{Nicolini2019Oct}, this finite potential can result in a singularity-free spacetime at $r = 0$ and modifies the metric surrounding an uncharged black hole.  It also plays the role of altering the entropy at the horizon and the Hawking temperature, which in turn affects the black hole's thermodynamic properties.

A interesting cosmological framework based on string T-duality and the incorporation of zero-point length within the gravitational potential was recently published by Luciano and Sheykhi \cite{Luciano2024Jul}.  They used the first law of thermodynamics applied to the apparent horizon of the FLRW Universe to arrive at the modified Friedmann equations.  By applying the modified Friedmann equations and using observational data to constrain the zero-point length around the Planck's scale, they studied the aspects of early Universe.  This complies with the T-duality of strings. Additionally, they investigated the evolution of density perturbation over the linear domain and discovered that the zero-point length significantly affects the growth of the density contrast.  The primordial seed fluctuations develop more slowly into large scale structure (LSS) as a result of this evolution of the density contrast.

Using the entropic force scenario and thermodynamics at the apparent horizon, Luciano and Sheykhi derived the modified Friedmann equations given by \cite{Luciano2024Jul} 
\begin{equation}
H^2+\frac{k}{a^2}-\alpha\left(H^2+\frac{k}{a^2}\right)^2+\mathcal{O}\left(\alpha^2\right)=\frac{8 \pi}{3} \rho_{tot} +\frac{\Lambda}{3},
\label{Fried_orig}
\end{equation}
where $\alpha = 3l_0^2/4$. Clearly, as the zero-point length $l_0 \to 0$, the original Friedmann equation in GR is recovered. The cosmological constant $\Lambda$ was assumed to be present in the energy-momentum tensor, which is given by $T_{\mu \nu} = (\rho_{tot} + p_{tot})u_\mu u_\nu + p_{tot} g_{\mu \nu}$. This tensor follows the continuity equation $\dot{\rho} + 3H(\rho_{tot} + p_{tot}) = 0$, where $H$ represents the Hubble parameter.  The total energy density and pressure are denoted by $\rho_{tot}$ and $p_{tot}$, respectively.  They employed cosmographic analysis to study the evolution of the Universe in different eras by using the modified Friedmann equation in a flat Universe setting ($k = 0$). 
In contrast with their work, we would assume that the Universe has a non-zero spatial curvature($k=\pm1$), as it has been shown to be very important in the very early universe, specially in the context of emergent Universe scenario \cite{ellis1,ellis2}. Moreover, in our setting we shall assume that topological defects contribute to the total energy density of the Universe along with a perfect fluid component obeying the equation of state (EoS) $p = \omega \rho$. Also, in this paper, we shall work with a vanishing cosmological constant, $\Lambda = 0$.

$n$-dimensional topological defects ($n = 0, 1, 2$) represent interesting possibilities in early-Universe cosmology. These defects may manifest themselves as a part of the total matter content of the Universe, represented by their unique EoS. In the context of early Universe cosmology, different types of topological defects find importance in various theoretical and observational aspects. Particularly, from the viewpoint of non-singular cosmology, Battista analyzed the impact of defects in the context of bouncing cosmology in GR \cite{Battista2021Aug}. 
In this present manuscript, we try to explore the non-singular emergent Universe filled with topological defects.

 Generically, a non-intercommuting network of $n$-dimensional topological defects can be thought of as an effective fluid with an EoS parameter $\omega_{n} = -n/3$ \cite{Faraoni2021Dec}. For example, monopoles, cosmic strings, and domain walls are some well-known topological defects corresponding to $n = 0, 1,$ and $2,$ respectively, yielding EoSs $(\omega_0, \omega_1, \omega_2) = (0, -1/3, -2/3)$. With respect to the topology of the associated symmetry groups, the topological defects may manifest as point-like, line-like, or surface-like shape, which are referred to as monopoles, strings and domain walls respectively. These topological defects exhibit stability, in a sense that monopoles are unlikely to decay into other particles, strings are unbreakable and domain walls cannot form holes which is ensured by topology and are also model independent. 

It may be noted that the energy density of the $n$-dimensional topological defect fluid evolves with the scale factor $a$ according to the relation \cite{Faraoni2021Dec}
\begin{equation}
\rho_D = \frac{\rho_D^{(0)}}{a^{3(1+\omega_n)}},
\label{rho_D}
\end{equation}
the energy density of the $n$-dimensional topological defects takes form 
\begin{equation}
\rho_D = \frac{\rho_D^{(0)}}{a^{3-n}},
\label{rho_D_n}
\end{equation}
by setting $\omega_n = -n/3$. As mentioned earlier, the total matter content is considered to be a combination of a non-interacting perfect fluid following a barotopic EoS and the effective fluid manifested by $n$-dimensional topological defect, we may write $\rho_{tot} = \rho + \rho_D$. Substituting this in Eq. \eqref{Fried_orig}, we obtain the first Friedmann equation which includes the contribution of the topological defect as
\begin{equation}
H^2 + \frac{k}{a^2} - \alpha \left(H^2 + \frac{k}{a^2}\right)^2 - \frac{C}{a^{3-n}} = \frac{8\pi \rho}{3},
\label{FE1}
\end{equation}
where $C = \frac{8\pi \rho_D^{(0)}}{3}$ is a constant. 
Differentiating Eq. \eqref{FE1} and using the conservation equation of the perfect fluid $\dot{\rho} + 3H(\rho + p) = 0$, one can derive the second Friedmann equation as
\begin{equation}
\left(\dot{H} - \frac{k}{a^2}\right)\left[1- 2\alpha\left(H^2 + \frac{k}{a^2}\right) \right] + \frac{C}{2a^{1-n}} = -4\pi (\rho + p).
\label{FE2}
\end{equation}
\section{Einstein static Universe and graceful exit mechanism}
\label{esu_eqs}
The basis of emergent cosmology is the idea that a stable ESU substitutes the initial singularity in the early Universe. The cornerstone of this framework is a set of fundamental requirements for a successful emergent scenario, which includes the requirement for a smooth and natural transition from a stable ESU. In this section, we shall thoroughly examine the stability of the ESU in the context of dynamical system analysis, stressing on the details of the phase transition mechanism that makes this exit possible.

In order to study the ESU, it is convenient to express the Raychaudhuri Eq. \eqref{FE2} in terms of the scale factor $a$. This gives
\begin{equation}
\ddot{a} = \frac{\dot{a}^2}{a} + \frac{k}{a} - \left[\frac{4\pi \rho a (1+\omega) + \frac{Ca^n}{2}}{1 - 2\alpha\left(\frac{\dot{a}^2}{a^2} + \frac{k}{a^2}\right)} \right],
\label{addot_gen}
\end{equation}
To discuss the ESU in the presence of topological defects, it will be convenient to consider all the types of topological defects separately. Before that, let us first calculate the energy density of the ESU $\rho_s$ by setting $\ddot{a} = \dot{a} = 0$ into Eq. \eqref{addot_gen}. This gives
\begin{equation}
\rho_s = \frac{C a_s^{n+3}+2k a_s^2+4 \alpha  k^2}{8 \pi  (\omega +1) a_s^3},
\label{rhos_gen}
\end{equation}
The existence condition of the ESU requires the positivity of $a_s, \alpha$ and $C$, which sets the constraint on $\omega$ as $\omega > -1$.

The stability of a dynamical system is performed based on the linearised system $\dot{x}_i = J_{ij} (x_j - x_{j0})$ around the critical point $(x_{1}, x_{2}) = (a_s, 0)$, where $a_s$ is the ES radius in our context. $J_{ij}$ represents the elements of the Jacobian $J$ defined as \cite{Gohain2024Jun}
\begin{equation}
\begin{aligned}
J = \left(\frac{\partial \mathcal{X}_i}{\partial x_j} \right)_{(a_{ES},0)} &= \begin{pmatrix}
\frac{\partial \mathcal{X}_1}{\partial x_1}\left.\right|_{(a_{ES},0)} & \frac{\partial \mathcal{X}_1}{\partial x_2}\left.\right|_{(a_{ES},0)} \\
\frac{\partial \mathcal{X}_2}{\partial x_1}\left.\right|_{(a_{ES},0)} & \frac{\partial \mathcal{X}_2}{\partial x_2}\left.\right|_{(a_{ES},0)}
\end{pmatrix} \\ &= \begin{pmatrix}
\hspace{-1cm} 0 & 1 \\
\frac{\partial \mathcal{X}_2}{\partial x_1}\left.\right|_{(a_{ES},0)} & 0
\end{pmatrix}.
\end{aligned}
\label{Jmat}
\end{equation}
The stability of the critical point $(x_1, x_2) = (a_{s}, 0)$ can be determined by the eigenvalues $\lambda$ of the $J$-matrix (\ref{Jmat}).
The eigenvalues of the Jacobian are obtained by calculating the roots of the equation
\begin{equation}
\lambda^2 - \lambda \, \mathrm{Tr} ( J ) + \mathrm{Det} (J) = 0,
\label{charac_eqn}
\end{equation}
The roots of Eq. \ref{charac_eqn} are
\begin{equation}
\lambda_{1,2} = \frac{1}{2} \left[ \mathrm{Tr} (J) \pm \sqrt{(\mathrm{Tr}(J))^2 - 4 \mathrm{Det} (J)} \right].
\label{eig_sol_gen}
\end{equation}
The stability of the critical point $(a_s, 0)$ can be interpreted from the sign of $\lambda^2$, i.e. for $\lambda^2 < 0$ and $\lambda^2 > 0$ the critical points are stable (center) and unstable (saddle) respectively.

Concerning the stability of the ESU under dynamical system approach and whether one can have a successful graceful exit or not, we may construct a autonomous system of differential equations by considering the following definition of variables
\begin{equation}
x_1 = a, \quad x_2 = \dot{a},
\label{variables}
\end{equation}
using which, from Eq. \eqref{addot_gen}
\begin{equation}
\begin{aligned}
&\dot{x}_1 = x_2,\\
&\dot{x}_2 = -\frac{x_1^2 \left(C x_1^n+8 \pi  \rho  x_1 (\omega +1)\right)}{2 \left(x_1^2-2 \alpha  \left(k+x_2^2\right)\right)}+\frac{k}{x_1}+\frac{x_2^2}{x_1}.
\end{aligned}
\label{dyn_gen}
\end{equation}
\subsection{A Closed Universe ($k=1$)}
Let us assume that the total matter content in a closed Universe ($k=1$) contain a perfect fluid with a barotopic EoS, obeying the relation $p = \omega \rho$ with n-dimensional topological defects. For the numerical analysis in this work, we need to choose the ES radius such that $a_s > l_p$ ($l_p$ is Planck's length). In this paper, we shall assume $a_s = 2$ for simplicity\footnote{Here, the Einstein static radius is a free constant, that can take any positive values provided that it is larger than the Planck's length. The value $a_s=2$ is chosen in such a way as to ensure it aligns with the classical design of emergent scenario. As we work in a natural unit system, $c = \hbar = 1$ and $G = 1$, the Planck length becomes $l_p = \sqrt{\frac{\hbar G}{c^3}} =1$. In order to bypass the quantum gravity regime, the ES radius should be larger than the Planck's length. Similarly, Planck's energy density should be is $\rho_{pl} = 1$. Thus the energy density of the ESU must be $0< \rho_s < 1$ to maintain conformity with a classical framework.}. This is in agreement with the setting that the ES radius can be arbitrarily positive but greater than the Planck length. This choice makes the calculations simpler to a great extent, as we shall see in the later sections.  Let us now systematically study the dynamical stability of the ESU and its transition to inflationary Universe for the following cases.
\subsubsection{Perfect fluid + Monopoles}
The first case we consider is a total fluid consisting of a perfect fluid and monopoles. In this regard, we set $k=1$ and $n = 0$ into Eq. \eqref{addot_gen}, which gives the energy density of the ESU as
\begin{equation}
\rho_s = \frac{4 \alpha +8 C+8}{64 \pi  (\omega +1)},
\label{rhos_k1_n0}
\end{equation}
Note that in this present work, we have already excluded the presence of the cosmological constant ($w = -1$), Therefore, the energy density given by the above equation (and in all other subsequent cases) does not blow up and always stays finite.

To generate the system of autonomous differential equations of interest we define the following variables
\begin{equation}
x_1 = a, \quad x_2 = \dot{a},
\label{variables1}
\end{equation}
using which, from Eq. \eqref{addot_gen} we get
\begin{equation}
\begin{aligned}
&\dot{x}_1 = x_2,\\
&\dot{x}_2 = -\frac{x_1^2 \left(C+8 \pi  \rho  x_1 (\omega +1)\right)}{2 \left(x_1^2-2 \alpha  \left(x_2^2+1\right)\right)}+\frac{x_2^2}{x_1}+\frac{1}{x_1}.
\end{aligned}
\label{dyn_k1_n0}
\end{equation}
The eigen value of the system \eqref{dyn_k1_n0} becomes
\begin{equation}
\lambda^2 = \frac{-8 \alpha ^2+32 \alpha +32 \alpha  C+32 \left(24 \pi  \alpha  (\omega +1) \rho _s-1\right)-512 \pi  (\omega +1) \rho _s}{2 (8-4 \alpha )^2},
\label{lamsq_k1_n0}
\end{equation}
For stable critical point $(a_s, 0)$, one must have $\lambda^2 < 0$. Using this condition we discuss the constraint on the EoS parameter $\omega$ as follows:
By setting $\lambda^2 < 0$ we obtain a critical value as
\[
\omega_{\mathrm{crit}}(\alpha,C,\rho_s) \;=\; \frac{4-4\alpha-4C\alpha+\alpha^2+64\pi\rho_s-96\pi\alpha\rho_s}{-64\pi\rho_s+96\pi\alpha\rho_s}\,.
\]
Then the existence region for a stable ESU can be written as the union of four regions: \footnote{The inequalities are derived using the \texttt{MATHEMATICA} software.}
\begin{equation}
\begin{array}{rcl}
\textbf{Case I:} &:& \text{ For } 0<\rho_s<1,\quad \alpha<\tfrac{2}{3},\quad C>0, \text{we must have  } \omega > \omega_{\mathrm{crit}}(\alpha,C,\rho_s),\\[1ex]
\textbf{Case II:} &:& \text{ For } 0<\rho_s<1,\quad \alpha=\tfrac{2}{3},\quad 0<C<\tfrac{2}{3}, \text{ no restriction on } \omega,\\[1ex]
\textbf{Case III:} &:& \text{ For } 0<\rho_s<1,\quad \tfrac{2}{3}<\alpha<2,\quad C>0, \text{ we must have } \omega < \omega_{\mathrm{crit}}(\alpha,C,\rho_s),\\[1ex]
\textbf{Case IV:} &:& \text{ For } 0<\rho_s<1,\quad \alpha>2,\quad C>0, \text{ we must have } \omega < \omega_{\mathrm{crit}}(\alpha,C,\rho_s)\,.
\end{array}
\label{omeg_exis_reg_lamsq}
\end{equation}

\begin{figure*}[htbp]
\centerline{\includegraphics[scale=0.5]{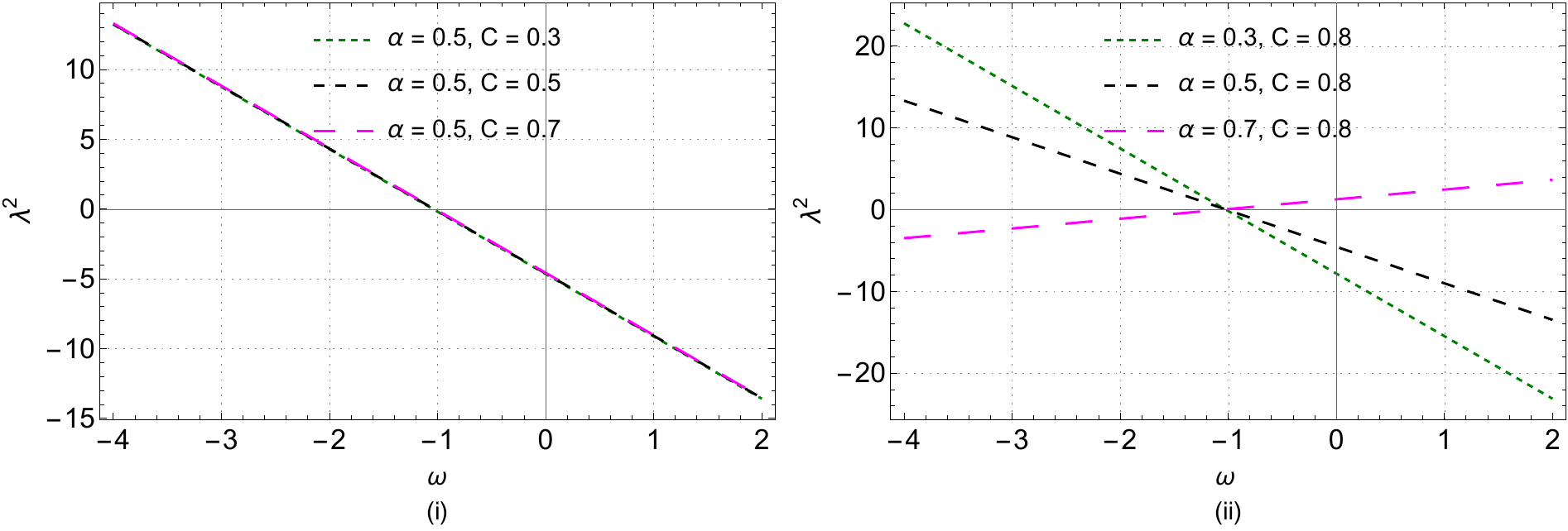}}
\caption{Phase transition of the ESU from stable ES phase to unstable inflationary phase is shown for $k = 1$ and $n = 0$ case.}
\label{lamsq_k1_n0_plt}
\end{figure*}
To view an explicit relation between the constants $\alpha$ and $c$ we need to eliminate $\rho_s$. Substituting Eq. \eqref{rhos_k1_n0} in inequation  \eqref{omeg_exis_reg_lamsq}
we obtain the existence region for a stable ESU as:
\begin{equation}
\text{Given } C > 0,\, \alpha \text{ must satisfy }\, 0 \le \alpha < -\frac{2}{5}\left(3+4C\right)+\frac{2\sqrt{2}}{5}\sqrt{12+17C+8C^2}\,.
\label{Exis_k1_n0_c_alp}
\end{equation}
It can be clearly seen from the Fig. \ref{lamsq_k1_n0_plt}, that the eigen value squared goes from the negative region to the positive region with decreasing value of the EoS parameter $\omega$\footnote{Intuitively, the evolution of $\omega$ from larger to smaller numerical values can thought of as cosmic time going from past to future. This is explained in detail in \cite{Khodadi2018Jun}.}. Thus, as time progresses the ESU evolves from a center-like critical point (negative $\lambda^2$) to a saddle like critical-point (positive $\lambda^2$). This implies that a phase transition occurs as the eigen-value squared evolves from negative to positive values representing the graceful exit from the stable ESU to the inflationary phase. For different combinations of the zero-point length parameter ($\alpha$) and the topological defect parameter $(C)$ (which in this case represents the monopole energy density), the evolution occurs in a similar pattern with no drastic variation in the $\lambda^2$ values with respect to the parameters.
\subsubsection{Perfect fluid + cosmic strings}
In the case of cosmic strings in a closed Universe, we set $k=1$ and $n = 1$ into Eq. \eqref{addot_gen}. This gives the energy density of the ESU as
\begin{equation}
\rho_s = \frac{4 \alpha +16 C+8}{64 \pi  (\omega +1)},
\label{rhos_k1_n1}
\end{equation}
In this case, we form the dynamical system as 
\begin{equation}
\begin{aligned}
&\dot{x}_1 = x_2,\\
&\dot{x}_2 = \frac{C x_1^3-8 \pi  \rho  x_1^3 (\omega +1)}{2 x_1^2-4 \alpha  \left(x_2^2+1\right)}+\frac{x_2^2}{x_1}+\frac{1}{x_1}.
\end{aligned}
\label{dyn_k1_n1}
\end{equation}
Solving for the eigen value squared, we obtain
\begin{equation}
\lambda^2 = \frac{8 \left(16 \pi  (\omega +1) \rho _s-2 C\right)}{(4 - 2 \alpha)^2}+\frac{6 C-48 \pi  (\omega +1) \rho _s}{4 - 2 \alpha}-\frac{1}{4},
\label{lamsq_k1_n1}
\end{equation}
For stable critical point $(a_s, 0)$, one must have $\lambda^2 < 0$, which sets the  constraints on $\omega$ just like the previous case by defining 
\[
\omega_{\mathrm{crit}}(\alpha,C,\rho_s) \;=\; \frac{4-8C-4\alpha+12C\alpha+\alpha^2+64\pi\rho_s-96\pi\alpha\rho_s}{-64\pi\rho_s+96\pi\alpha\rho_s}\,.
\]
Then the existence regions condition can be interpreted as the union of four regions:
\begin{equation}
\begin{array}{rcl}
\textbf{Case I:} &:& \text{ For } 0<\alpha<\frac{2}{3},\quad 0<\rho_s<1,\quad C>0, \text{ we must  have  } \omega>\omega_{\mathrm{crit}}(\alpha,C,\rho_s),\\[1ex]
\textbf{Case II:} &:& \text{ For } \alpha=\frac{2}{3},\quad 0<\rho_s<1,\quad C>0, \text{ no restriction on } \omega, \\[1ex]
\textbf{Case III:} &:& \text{ For } \frac{2}{3}<\alpha<2,\quad 0<\rho_s<1,\quad C>0,\text{ we must  have  } \omega<\omega_{\mathrm{crit}}(\alpha,C,\rho_s),\\[1ex]
\textbf{Case IV:} &:& \text{ For } \alpha>2,\quad 0<\rho_s<1,\quad C>0,\text{ we must  have  } \omega<\omega_{\mathrm{crit}}(\alpha,C,\rho_s)\,.
\end{array}
\label{omeg_exis_reg_lamsq_cs}
\end{equation}

\begin{figure*}[htbp]
\centerline{\includegraphics[scale=0.5]{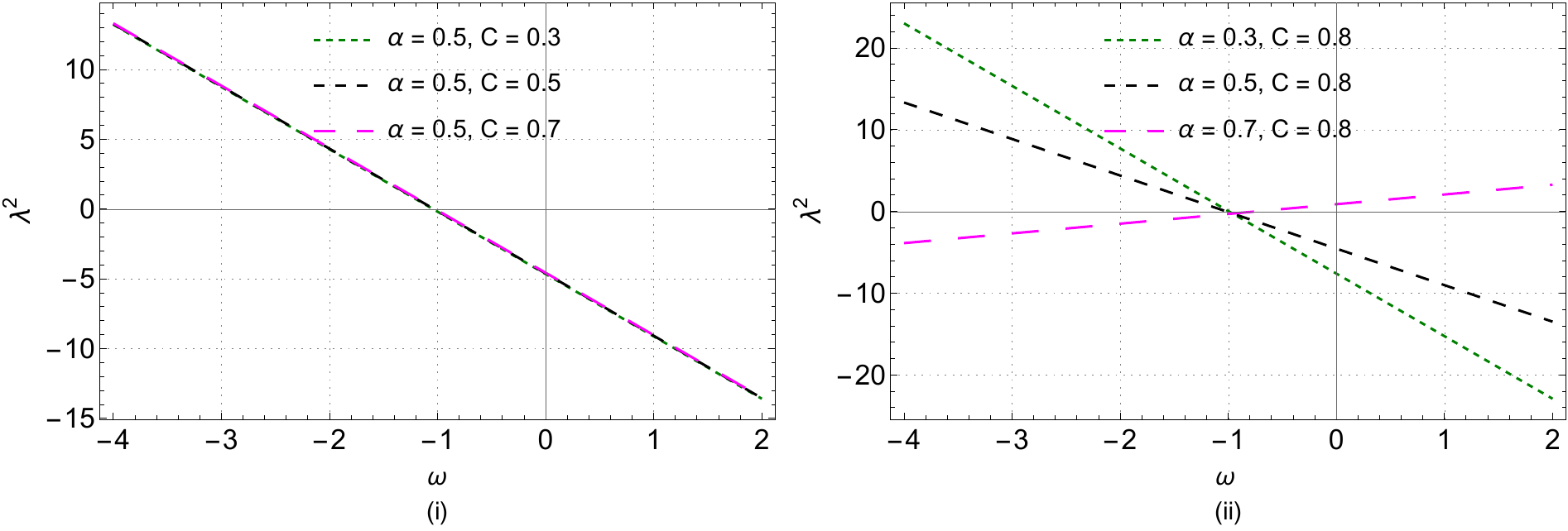}}
\caption{Phase transition of the ESU from stable ES phase to unstable inflationary phase is shown for $ k = 1$ and $n = 1$ case.}
\label{lamsq_k1_n1_plt}
\end{figure*}
Again to find an explicit relation between the constants $\alpha$ and $c$ we substitute Eq. \eqref{rhos_k1_n1} in the inequation \eqref{omeg_exis_reg_lamsq_cs}
to obtain the existence region for a stable ESU as:
\begin{equation}
\text{Given }C > 0,\, \alpha \text{ should satisfy } 0 \le \alpha < -\frac{6}{5}(1 + C) + \frac{2}{5} \sqrt{24 + 28 C + 9 C^2}
\label{Exis_k1_n1_c_alp}
\end{equation}

\subsubsection{Perfect fluid + Domain walls}
In the presence of domain walls in a closed Universe, we set $k=1$ and $n = 2$ into Eq. \eqref{addot_gen}. This gives the energy density of the ESU as
\begin{equation}
\rho_s = \frac{4 \alpha +32 C+8}{64 \pi  (\omega +1)},
\label{rhos_k1_n2}
\end{equation}
Proceeding in a similar way as in the earlier cases, we form the autonomous system of equations as 
\begin{equation}
\begin{aligned}
&\dot{x}_1 = x_2,\\
&\dot{x}_2 = \frac{C x_1^4-8 \pi  \rho  x_1^3 (\omega +1)}{2 x_1^2-4 \alpha  \left(x_2^2+1\right)}+\frac{x_2^2}{x_1}+\frac{1}{x_1}.
\end{aligned}
\label{dyn_k1_n2}
\end{equation}
Now, solving for the eigen value squared, we get
\begin{equation}
\lambda^2 = \frac{8 \left(16 \pi  (\omega +1) \rho _s-4 C\right)}{(4-2 \alpha )^2}+\frac{16 C-48 \pi  (\omega +1) \rho _s}{4-2 \alpha }-\frac{1}{4},
\label{lamsq_k1_n2}
\end{equation}
For stable critical point $(a_s, 0)$, one requires $\lambda^2 < 0$, which leads to $\omega_{\mathrm{crit}}$ as \[
\omega_{\mathrm{crit}}(\alpha,C,\rho_s) \;=\; 
\frac{4-32C-4\alpha+32C\alpha+\alpha^2+64\pi\rho_s-96\pi\alpha\rho_s}{-64\pi\rho_s+96\pi\alpha\rho_s}\,.
\]
Then the existence condition can be expressed as union of four regions:
\begin{equation}
\begin{array}{rcl}
\textbf{Case I:} &:& \text{ For } 0<\alpha<\tfrac{2}{3},\quad 0<\rho_s<1,\quad C>0,\text{ we have } 
\omega > \omega_{\mathrm{crit}}(\alpha,C,\rho_s),\\[1ex]
\textbf{Case II:} &:& \text{ For } \alpha=\tfrac{2}{3},\quad 0<\rho_s<1,\quad 0<C<\tfrac{1}{6},\text{ no restriction on } \omega \\[1ex]
\textbf{Case III:} &:& \text{ For } \tfrac{2}{3}<\alpha<2,\quad 0<\rho_s<1,\quad C>0,\text{ we have }
\omega < \omega_{\mathrm{crit}}(\alpha,C,\rho_s),\\[1ex]
\textbf{Case IV:} &:& \text{ For } \alpha>2,\quad 0<\rho_s<1,\quad C>0, \text{ we have }
\omega < \omega_{\mathrm{crit}}(\alpha,C,\rho_s)\,.
\end{array}
\label{omeg_exis_reg_lamsq_dw}
\end{equation}


\begin{figure*}[htbp]
\centerline{\includegraphics[scale=0.5]{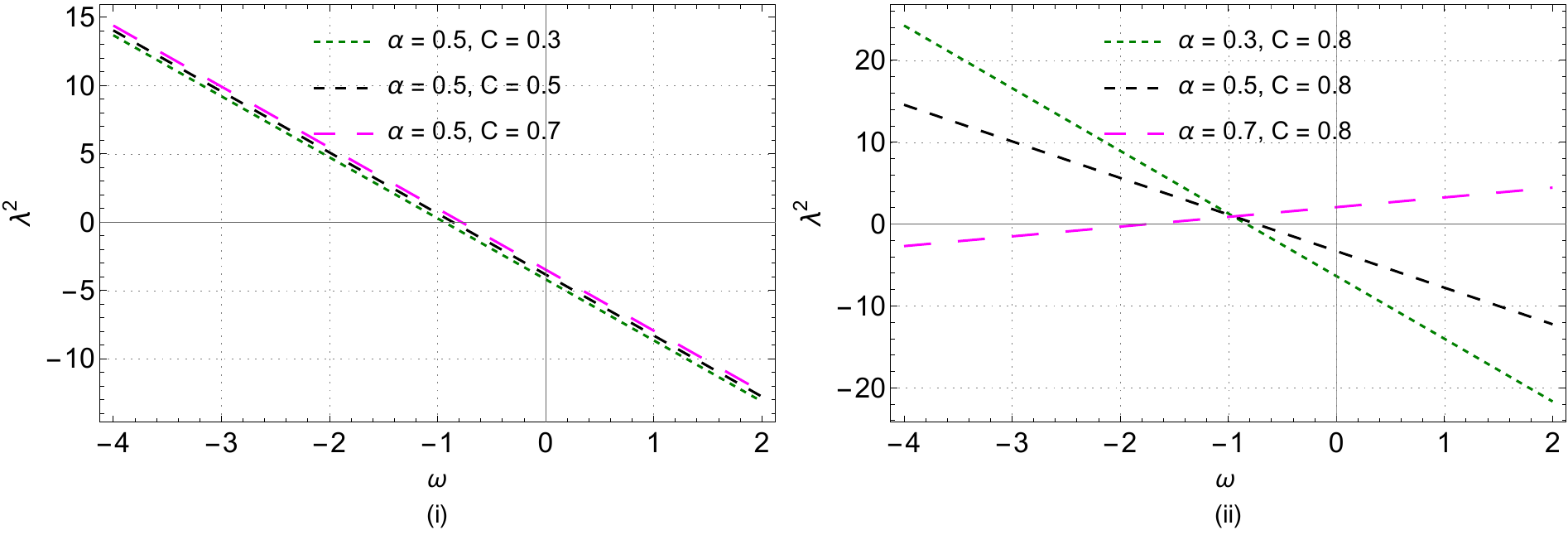}}
\caption{Phase transition of the ESU from stable ES phase to unstable inflationary phase is shown for $k = 1$ and $n = 2$ case.}
\label{lamsq_k1_n2_plt}
\end{figure*}
Following the same procedure as before, to find an explicit relation between the constants $\alpha$ and $c$ we substitute Eq. \eqref{rhos_k1_n2} in inquation \eqref{omeg_exis_reg_lamsq_dw}
to obtain the existence region for a stable ESU as
\begin{equation}
\text{Given } C > 0, \alpha \text{ must lie within } 
0 \le \alpha < -\frac{2}{5}(3+4C) + \frac{4}{5}\sqrt{2}\sqrt{3+3C+2C^2}\,.
\label{Exis_k1_n2_c_alp}
\end{equation}
Note that in the case of closed Universe, different values of $C$, for a fixed value of $\alpha$ does not noticeably influence the variation of $\lambda^2$. However, the variation of $\lambda^2$ is significantly sensitive to the values of $\alpha$. For instance, $\alpha$ higher than 0.5, $\lambda^2$ goes from positive to negative values as $\omega$ goes from positive to negative (or as time progresses). This indicates there is no successful occurrence of graceful exit from the ES phase to inflation.
\begin{figure*}[htb]
\centerline{\includegraphics[scale=0.45]{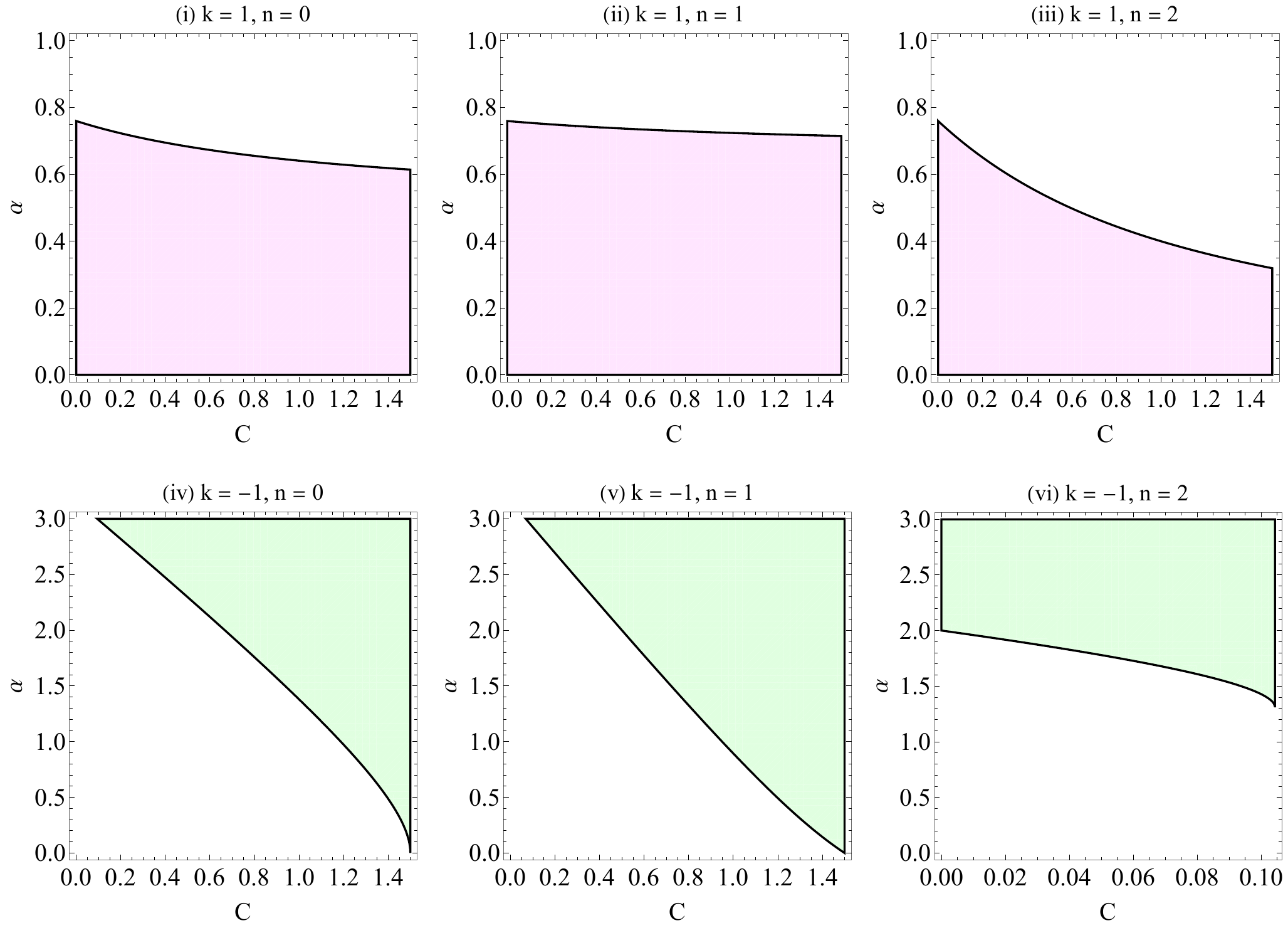}}
\caption{The parameter space for stable ESU is shown for all dimensions of topological defects for both $k = 1$ and $k = -1$ cases.}
\label{region_plots}
\end{figure*}
The parameter space denoting the possible values for stable ESU for closed Universe, for all three dimensions of topological defects is shown in Fig. \ref{region_plots}.

\subsection{An Open Universe ($k=-1$)}
Let us now assume that the total matter content in a open Universe ($k=-1$) contain a perfect fluid with a barotopic EoS, obeying the relation $p = \omega \rho$ with n-dimensional topological defects. In this regard, let us study the following possibilities:
\subsubsection{Perfect fluid + Monopoles}
In this case, we set $k=-1$ and $n=0$ into Eq. \eqref{addot_gen}, which gives the energy density of the ESU as
\begin{equation}
\rho_s = \frac{4 \alpha +8 C-8}{64 \pi  (\omega +1)},
\label{rhos_k-1_n0}
\end{equation}
Again by using the definition
\begin{equation}
x_1 = a, \quad x_2 = \dot{a},
\label{variablesk-1}
\end{equation}
we construct the dynamical system from Eq. \eqref{addot_gen} as
\begin{equation}
\begin{aligned}
&\dot{x}_1 = x_2,\\
&\dot{x}_2 = \frac{C x_1^2-8 \pi  \rho  x_1^3 (\omega +1)}{2 x_1^2-4 \alpha  \left(x_2^2-1\right)}+\frac{x_2^2}{x_1}-\frac{1}{x_1}.
\end{aligned}
\label{dyn_k-1_n0}
\end{equation}
The eigenvalue of the system \eqref{dyn_k-1_n0} becomes
\begin{equation}
\lambda^2 = \frac{8 \left(16 \pi  (\omega +1) \rho _s-C\right)}{(2 \alpha +4)^2}+\frac{2 C-48 \pi  (\omega +1) \rho _s}{2 \alpha +4}+\frac{1}{4},
\label{lamsq_k-1_n0}
\end{equation}
For stable critical point $(a_s, 0)$, one must have $\lambda^2 < 0$, which sets the following constraint on $\omega$ as follows:
\begin{equation}
\text{Given, } 0 < \rho_s < 1,\, \alpha > 0\, \text{ and } C > 0  \,\text{ we should have } \,
\omega > \frac{4 + 4\alpha + 4C\alpha + \alpha^2 - 64\pi\rho_s - 96\pi\alpha\rho_s}{64\pi\rho_s + 96\pi\alpha\rho_s}.
\label{omeg_exis_reg_lamsq_k-1_n0}
\end{equation}
\begin{figure*}[htbp]
\centerline{\includegraphics[scale=0.5]{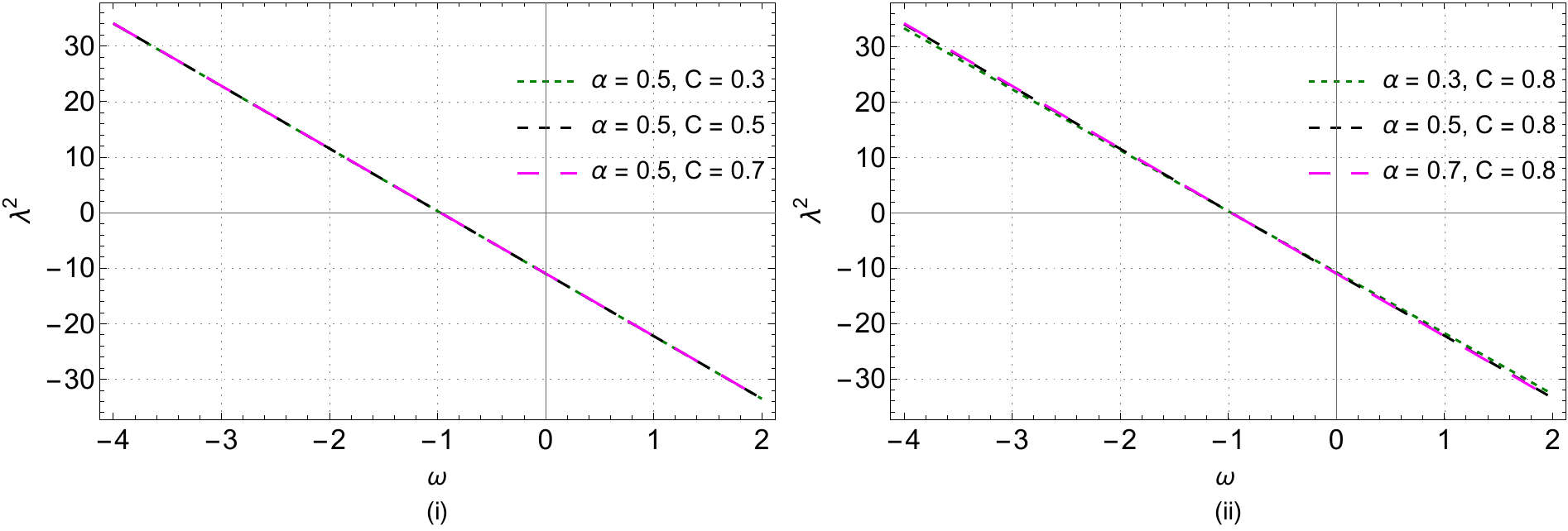}}
\caption{Phase transition of the ESU from stable ES phase to unstable inflationary phase is shown for $ k = -1$ and $n = 0$ case.}
\label{lamsq_k-1_n0_plt}
\end{figure*}
To view an explicit relation between the constants $\alpha$ and $c$ we need to eliminate $\rho_s$. Substituting Eq. \eqref{rhos_k-1_n0} into inequation \eqref{omeg_exis_reg_lamsq_k-1_n0}
we obtain the existence region for a stable ESU. $\text{Given,} \, 0 < C < \frac{3}{2} \text{ we must have } $
\begin{equation}
\quad \alpha > \frac{2}{5}(3-2C) + \frac{2\sqrt{2}}{5}\sqrt{12-11C+2C^2}\,.
\label{Exis_k-1_n0_c_alp}
\end{equation}
\subsubsection{Perfect fluid + cosmic strings}
In this case of cosmic strings, we set $k=-1$ and $n = 1$ into Eq. \eqref{addot_gen}. This gives the energy density of the ESU as
\begin{equation}
\rho_s = \frac{\alpha +4 C-2}{16 \pi  (\omega +1)},
\label{rhos_k-1_n1}
\end{equation}
The autonomous system of equations can be formulated as 
\begin{equation}
\begin{aligned}
&\dot{x}_1 = x_2,\\
&\dot{x}_2 = \frac{C x_1^{n+2}-8 \pi  \rho  x_1^3 (\omega +1)}{2 x_1^2-4 \alpha  \left(x_2^2-1\right)}+\frac{x_2^2}{x_1}-\frac{1}{x_1}.
\end{aligned}
\label{dyn_k-1_n1}
\end{equation}
The eigen value squared for this case is
\begin{equation}
\lambda^2 = \frac{8 \left(16 \pi  (\omega +1) \rho _s-2 C\right)}{(4+2 \alpha )^2}+\frac{6 C-48 \pi  (\omega +1) \rho _s}{4+2 \alpha }+\frac{1}{4},
\label{lamsq_k-1_n1}
\end{equation}
For stable critical point $(a_s, 0)$, one must have $\lambda^2 < 0$, which sets the following constraint on $\omega$ as:
\begin{equation}
\text{Given, } 0 < \rho_s < 1, \, \alpha > 0, \, C > 0 \text{ one must have } \, 
\omega > \frac{4 + 8C + 4\alpha + 12C\alpha + \alpha^2 - 64\pi \rho_s - 96\pi\alpha\rho_s}
{64\pi \rho_s + 96\pi\alpha\rho_s}.
\label{omeg_exis_reg_lamsq_k-1_n1}
\end{equation}
\begin{figure*}[htbp]
\centerline{\includegraphics[scale=0.5]{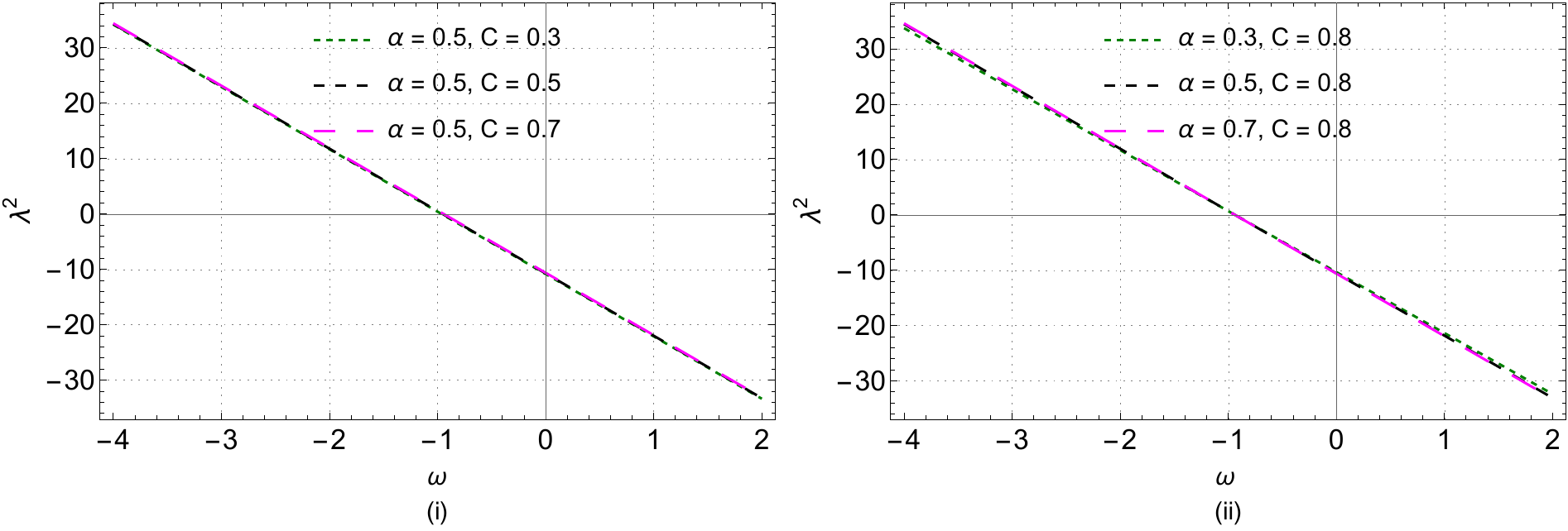}}
\caption{Phase transition of the ESU from stable ES phase to unstable inflationary phase is shown for $k = -1$ and $n = 1$ case.}
\label{lamsq_k-1_n1_plt}
\end{figure*}
To find an explicit relation between the constants $\alpha$ and $C$, we substitute Eq. \eqref{rhos_k-1_n1} in inequation \eqref{omeg_exis_reg_lamsq_k-1_n1}
to obtain the existence region for a stable ESU as. In this case, Given \, $0 < C < 3/2$,  we must have 
\begin{equation}
 \alpha > -\frac{6}{5}(-1+C) + \frac{2}{5}\sqrt{24-28C+9C^2}
\label{Exis_k-1_n1_c_alp}
\end{equation}
\subsubsection{Perfect fluid + Domain walls}
In the presence of domain walls in a open Universe, we set $k=-1$ and $n = 2$ into Eq. \eqref{addot_gen}. This gives the energy density of the ESU as
\begin{equation}
\rho_s = \frac{\alpha +8 C-2}{16 \pi  (\omega +1)},
\label{rhos_k-1_n2}
\end{equation}
The autonomous system of equations in this case is 
\begin{equation}
\begin{aligned}
&\dot{x}_1 = x_2,\\
&\dot{x}_2 = \frac{C x_1^4-8 \pi  \rho  x_1^3 (\omega +1)}{2 \left(2 \alpha  \left(-x_2^2-1\right)+x_1^2\right)}+\frac{x_2^2}{x_1}-\frac{1}{x_1}.
\end{aligned}
\label{dyn_k-1_n2}
\end{equation}
Solving for the eigen value squared we get
\begin{equation}
\lambda^2 = \frac{8 \left(16 \pi  (\omega +1) \rho _s-4 C\right)}{(4-2 \alpha )^2}+\frac{16 C-48 \pi  (\omega +1) \rho _s}{4-2 \alpha }+\frac{1}{4},
\label{lamsq_k-1_n2}
\end{equation}
For stable critical point $(a_s, 0)$, setting $\lambda^2 < 0$ leads to the following constraint on $\omega$. Provided $ 0<\rho_s<1$ and  $C>0$, let us define a critical value of $\omega$ as:
\begin{equation}
\begin{aligned}
&\quad \omega_{crit}(\alpha,C,\rho_s)=\frac{-4-32C+4\alpha+32C\alpha-\alpha^2+64\pi\rho_s-96\pi\alpha\rho_s}{-64\pi\rho_s+96\pi\alpha\rho_s}.\\[1mm]
&\hspace{-3.8cm}\text{Then, the existence condition can be given as:}\\[1mm]
&\quad
\text{For } \begin{cases}
0<\alpha<\tfrac{2}{3}: \quad \omega>\omega_{crit}(\alpha,C,\rho_s),\\[1mm]
\alpha>\tfrac{2}{3}: \quad \omega<\omega_{crit}(\alpha,C,\rho_s).
\end{cases}
\end{aligned}
\label{omeg_exis_reg_lamsq_k-1dw}
\end{equation}
\begin{figure*}[htbp]
\centerline{\includegraphics[scale=0.5]{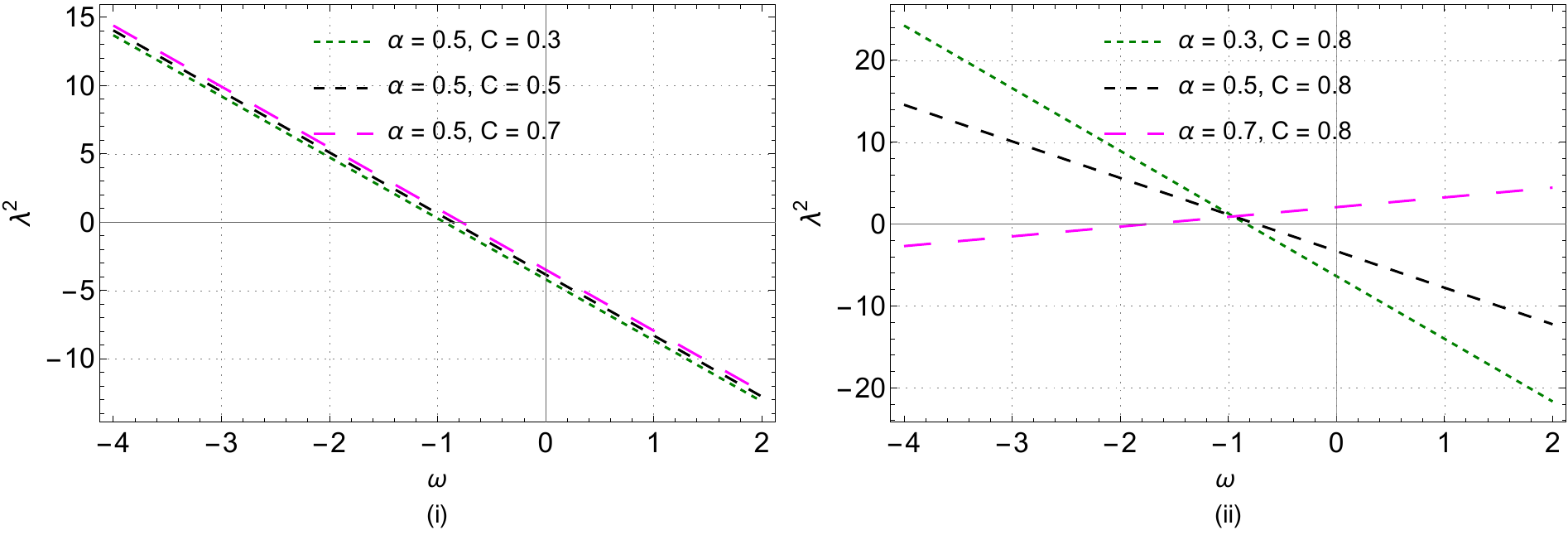}}
\caption{Phase transition of the ESU from stable ES phase to unstable inflationary phase is shown for $ k = -1$ and $n = 2$ case.}
\label{lamsq_k-1_n2_plt}
\end{figure*}
Following the same procedure as before, to find an explicit relation between the constants $\alpha$ and $C$ we substitute Eq. \eqref{rhos_k-1_n2} in inequation \eqref{omeg_exis_reg_lamsq_k-1dw}
to obtain the existence region for a stable ESU. Given,\, $0 < C < \frac{1}{4}\Bigl(5-\sqrt{21}\Bigr)$ we must have
\begin{equation}
\alpha > \frac{2}{7}(5-4C) + \frac{4}{7}\sqrt{\,1-10C+4C^2}\,.
\label{Exis_k-1_n2_c_alp}
\end{equation}
In the case of open Universe, we find that the occurrence of graceful exit is not affected drastically by the model parameters, with no observable constraint on $\alpha$ and $C$ unlike the case of closed Universe.
The parameter space denoting the possible values for stable ESU for open Universe for all kinds of topological defects is shown in Fig. \ref{region_plots}.

\section{Stability under Homogeneous Linear Perturbation}
\label{hom_scalar_pert}
Our goal in this section is to understand the stability of the ES Universe for $k = \pm 1$ Universes under linear homogeneous perturbations. We want to determine whether such perturbations could have an impact on the ESU's stability.  A time-dependent perturbation is added to the energy density and scale factor upto a linear order given by
\begin{equation}
a(t) = a_s(1 + \delta a(t)), \quad \rho(t) = \rho_s (1+ \delta \rho(t)),
\label{pert1}
\end{equation}
where $\delta a(t)$ and $\delta \rho(t)$ are infinitesimal linear perturbations introduced to the scale factor and energy density respectively.

For the ESU, setting $\dot{a} = \ddot{a} = 0$ in equation \eqref{FE1} and \eqref{FE2} we obtain
\begin{equation}
\rho_s d\rho = \frac{3}{8\pi}\left(\frac{4\alpha k^2}{a_s^4} - \frac{2k}{a_s^4} - \frac{C(3-n)}{a_s^{3-n}}\right)\delta a,
\label{rhodrho1}
\end{equation}
\begin{equation}
 -\frac{k}{a_s^2} + \frac{2\alpha k^2}{a_s^4} + \frac{C}{2a_s^{1-n}}+ 4\pi\rho_s (1+\omega) = 0.
\label{perteq1}
\end{equation}
Now, inserting the perturbed scale factor and energy density given by Eqs \eqref{pert1} in equation \eqref{FE2}, and using Eq. \eqref{rhodrho1} and \eqref{perteq1}, we finally obtain 
\begin{equation}
\delta \ddot{a} + \Omega^2 \, \delta a  = 0,
\label{perteq3}
\end{equation}
where
\begin{equation}
\begin{aligned}
\Omega^2 = \frac{1}{2 a_s^2 \left(a_s^2-2 \alpha  k\right)}&\left[3 C (n-3) (\omega +1) a_s^{n+1}+C (n-1) a_s^{n+3} \right.\\& \left. -2 a_s^2 (3 k \omega +k)+4 \alpha  k^2 (3 \omega +7)\right],
\end{aligned}
\label{Omeg1}
\end{equation}
For finite oscillating perturbation modes, which admit stable ES solutions one must have, $\Omega > 0$. The stable solutions of the equations \eqref{perteq3} is of the form
\begin{equation}
\delta a (t) = C_1 e^{i \Omega t} + C_2 e^{-i\Omega t},
\label{sol1}
\end{equation}
where $C_1$ and $C_2$ are integration constants. 
Therefore, given the condition $\Omega > 0$, the stability intervals are obtained for both closed and open Universes as follows:
\begin{figure*}[tbh]
\centerline
\centerline{\includegraphics[scale=0.4]{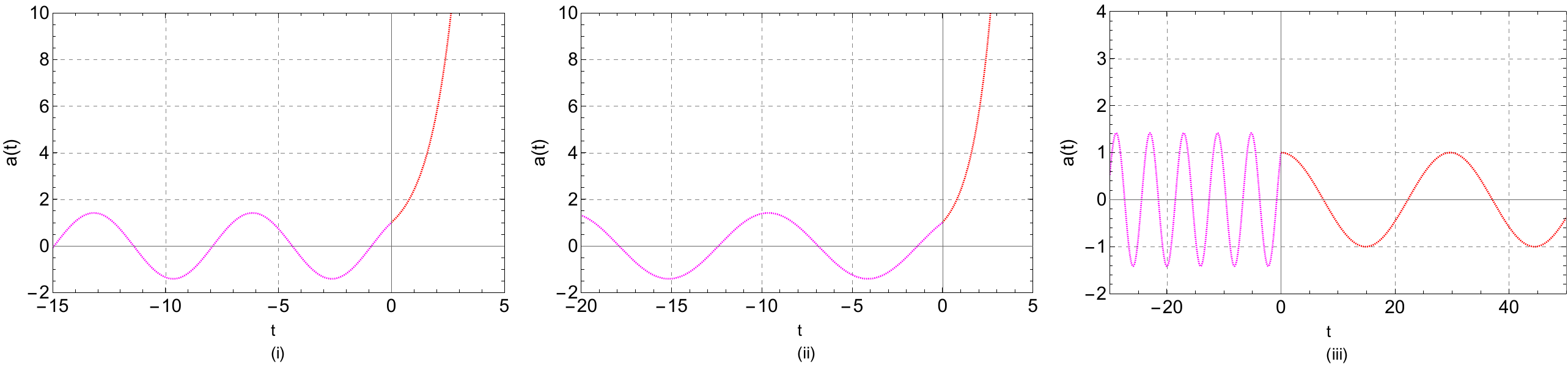}}
\caption{Graceful exit mechanism for $k=1$ under the influence of homogeneous perturbation is shown. The oscillations represented in magenta occurs when $\omega = 0.3$, $\alpha = 0.5$ and $C = 0.6$. The red exponential curve (in (i) and (ii) corresponding to monopoles and cosmic strings) occurs when $\omega$ is switched to $-0.5$. Whereas in (iii) for domain walls, inspite to this switching, the oscillations does not diverge and maintains the preceding pattern.}
\label{Grace_homo1}
\end{figure*}
\begin{figure*}[tbh]
\centerline
\centerline{\includegraphics[scale=0.4]{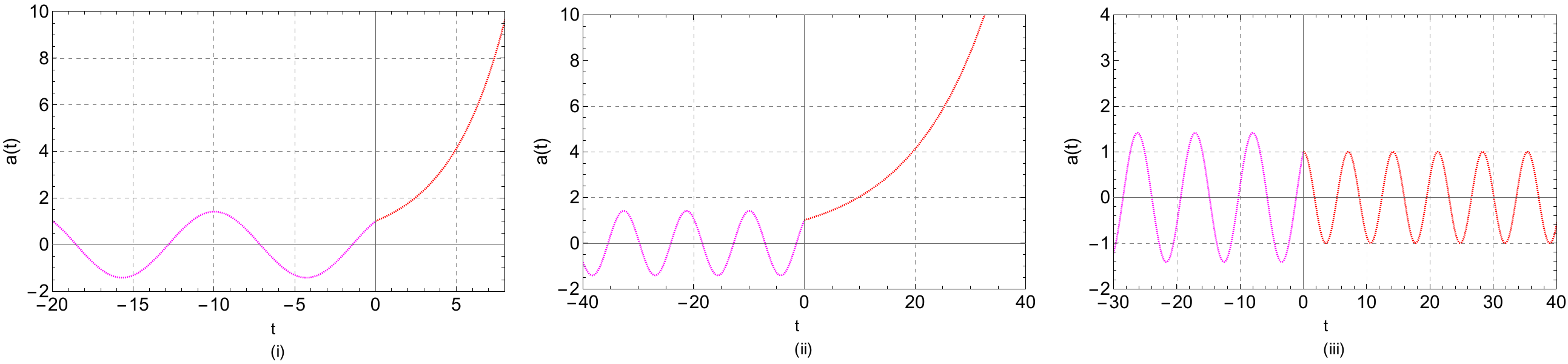}}
\caption{Graceful exit mechanism for $k=-1$ under the influence of homogeneous perturbation is shown. The oscillations represented in magenta in this case occurs when $\omega = -0.5$, $\alpha = 0.5$ and $C = 0.6$. The red exponential curve (in (i) and (ii) corresponding to monopoles and cosmic strings) occurs when $\omega$ is switched to $0.3$. Whereas in (iii) for domain walls, like the previous case, inspite to this switching of the values of $\omega$, the oscillations does not diverge as well.}
\label{Grace_homo2}
\end{figure*}
Like the previous analysis, we have set $a_s = 2$ for simplicity in this case also. 
The graceful exit mechanism from a stable ES phase to the inflationary phase under the effect of scalar homogeneous perturbation is depicted in Figs. \ref{Grace_homo1} and \ref{Grace_homo2}. The ESU to inflation phase transition point in the current analysis is defined at $t_0 = 0$. For plotting purposes we have set the integration constants to arbitrary positive values, i.e. $C_1 = C_2 = 1$. Setting $\omega = 0.3,\alpha = 0.5,C = 0.6$ yields the stable region of the ESU subjected to homogeneous perturbation in the case of a closed universe. The exponential inflationary phase is initiated when the value of $\omega$ changes from $0.3$ to $-0.5$, by breaking the scale factor's infinite loop of oscillations. But, for $\omega > 0$, the oscillations do not break and thus a graceful exit does not occur. Therefore, a fluid obeying the strong energy condition (SEC) does not allow a successful graceful exit from the ESU in a closed Universe in the presence of monopoles and cosmic strings (see Figs. \ref{Grace_homo1} (i) and (ii)). Violating the SEC may typically imply that the pressure of the perfect fluid is negative, which can drive the accelerated expansion needed for inflation. In contrast, the presence of domain walls causes the finite sinusoidal oscillations to stay unbroken, preventing a graceful exit into the inflationary phase regardless of the perfect fluid's EoS (see Fig. \ref{Grace_homo1} (iii)). Now for the open Universe case, by setting $\omega = -0.5,\alpha = 0.5,C = 0.6$, we find that ESU is stable subject to homogeneous perturbation, but breaks when $\omega = 0.3$, in the presence of monopoles and cosmic strings which is opposite to the aforementioned observation made in the context of the closed Universe. Thus, a fluid violating the SEC does not allow a successful graceful exit from the ESU in an open Universe in the presence of monopoles and cosmic strings (see Figs. \ref{Grace_homo2} (i) and (ii)). Therefore, attaining a graceful exit in an open Universe does not strictly necessitate exotic physics but can be accomplished with standard cosmological fluids that obey the SEC. However, on a similar note to the closed Universe case, one does not observe a graceful exit whatsoever in the presence of domain walls in an open Universe (see Fig. \ref{Grace_homo2} (iii)). In a nutshell, under scalar homogeneous perturbation, a successful graceful exit from ESU to the standard inflationary cosmology demands the requirement of a fluid that violates the SEC ($\omega < 0$) for an closed Universe whereas requires a perfect fluid that obeys the SEC for open Universe case.

Let us now discuss the stability intervals for stable ESU under homogeneous perturbation for both $k = 1$ and $-1$. To do this, let us first define a critical value of \(\omega\) for a clear understanding. This critical value arises from the condition for finite oscillation modes. 

\bigskip
\noindent\textbf{Closed Universe (\(k=1\))}

\medskip
\noindent\textbf{Monopoles (\(n=0\)):} Let us define 
\[
\omega_{\mathrm{crit}}^{(0)} = \frac{-4 - 13C + 14\alpha}{12 + 9C - 6\alpha}.
\]
Then the stability conditions are met if one the following holds:
\(
\text{ Provided }  C > 0, 
 \text{ (i) }  0 < \alpha < 2 \text{ and } \omega < \omega_{\mathrm{crit}}^{(0)}, \text{  or  }  
\text{(ii) }  2 < \alpha < \tfrac{1}{2}(4+3C) \text{ and } \omega > \omega_{\mathrm{crit}}^{(0)}, \text{ or } 
\text{(iii) }\alpha > \tfrac{1}{2}(4+3C) \text{ and }  \omega < \omega_{\mathrm{crit}}^{(0)}.
\)

\medskip
\noindent\textbf{Cosmic Strings (\(n=1\)):} In this case let us define 
\[
\omega_{\mathrm{crit}}^{(1)} = \frac{-2 - 6C + 7\alpha}{6 + 6C - 3\alpha}.
\]
The stability conditions are:
\(
\text{Provided }  C > 0, 
\text{ (i) } 0 < \alpha < 2 \text{ and } \omega < \omega_{\mathrm{crit}}^{(1)},
\text{ (ii) }  2 < \alpha < 2+2C \text{ and }\omega > \omega_{\mathrm{crit}}^{(1)},
\text{ (iii) } \alpha > 2+2C \text{ and } \omega < \omega_{\mathrm{crit}}^{(1)}.
\)

\medskip
\noindent\textbf{Domain Walls (\(n=2\)):} Let us again define 
\[
\omega_{\mathrm{crit}}^{(2)} = \frac{-2 + 2C + 7\alpha}{6 + 6C - 3\alpha}.
\]
The stability conditions to be satisfied are:
\(
\text{Provided } C > 0, 
\text{ (i) } 0 < \alpha < 2 \text{ and } \omega < \omega_{\mathrm{crit}}^{(2)}, 
\text{ (ii) } 2 < \alpha < 2+2C \text{ and } \omega > \omega_{\mathrm{crit}}^{(2)},
\text{ (iii) } \alpha > 2+2C \text{ and } \omega < \omega_{\mathrm{crit}}^{(2)}.
\)

\bigskip
\noindent\textbf{Open Universe (\(k=-1\))}

In the case of open universe, the critical values of $\omega$ are modified. Let us introduce a parameter \(C^*\) that depends on \(\alpha\).
Thus the stability conditions for all topological defects are found as

\medskip
\noindent\textbf{For Monopoles (\(n=0\)):} Define 
\[
\omega_{\mathrm{crit}}^{(0,op)} = \frac{4 - 13C + 14\alpha}{-12 + 9C - 6\alpha} \quad \text{and} \quad C^* = \frac{1}{3}(4+2\alpha).
\]
For \(\alpha > 0\), the stability conditions are:
\(
\text{If } C \le C^*, \quad  \omega > \omega_{\mathrm{crit}}^{(0,op)}, \quad
\text{ and if }  \quad C > C^*, \quad  \omega < \omega_{\mathrm{crit}}^{(0,op)}.
\)

\medskip
\noindent\textbf{For Cosmic Strings (\(n=1\)):} Define 
\[
\omega_{\mathrm{crit}}^{(1,op)} = \frac{2 - 6C + 7\alpha}{-6 + 6C - 3\alpha} \quad \text{and} \quad C^* = \frac{2+\alpha}{2}.
\]
For \(\alpha > 0\):
\(
\text{If } C \le C^*, \quad  \omega > \omega_{\mathrm{crit}}^{(1,op)}, \quad 
\text{ and if } \quad  C > C^*, \quad  \omega < \omega_{\mathrm{crit}}^{(1,op)}.
\)

\medskip
\noindent\textbf{For Domain Walls (\(n=2\)):} Define 
\[
\omega_{\mathrm{crit}}^{(2,op)} = \frac{2+2C+7\alpha}{-6+6C-3\alpha} \quad \text{and} \quad C^* = \frac{2+\alpha}{2}.
\]
For \(\alpha > 0\):
\(
\text{If } C \le C^*, \quad  \omega > \omega_{\mathrm{crit}}^{(2,op)}, \quad 
\text{ and if } \quad  C > C^*, \quad  \omega < \omega_{\mathrm{crit}}^{(2,op)}.
\)

\section{Stability under Inhomogeneous Density Perturbation}
\label{inhomo_dens}
The evolution equation of inhomogeneous density perturbation in an FLRW background is given by (in the absence of $\Lambda$) \cite{Barrow2003May}
\begin{equation}
\ddot{\Delta}+\left(2-6 \omega+3 c_{\mathrm{s}}^2\right) H \dot{\Delta}+\left[12\left(\omega-c_{\mathrm{s}}^2\right) \frac{k}{a^2} +4 \pi G\left(3 \omega^2+6 c_{\mathrm{s}}^2-8 \omega-1\right) \rho\right] \Delta  -c_{\mathrm{s}}^2 \mathrm{D}^2 \Delta- \omega\left(\mathrm{D}^2+3 \frac{k}{a^2}\right) \mathcal{E}=0.
\label{pert_dens_eq}
\end{equation}
The inhomogeneous density perturbations are defined using the 1+3 - covariant gauge-invariant approach as $\Delta = a^2 \mathrm{D}^2 \rho / \rho$, with $\mathrm{D}$ as the covariant spatial Laplacian. Here $c_s^2 = dp/d\rho$ and $\mathcal{E} = (a^2 \mathrm{D}^2 p - \rho c_s^2 \Delta)/p$ are the squared-sound speed and entropy perturbation respectively. 

For the closed Universe ($k = 1$) case, after setting the conditions for ESU ($H = 0$) and also the fact that for an ESU $\mathcal{E} = 0$ and without and as previously mentioned, conveniently choosing $a_s = 2$, the Eq. \eqref{pert_dens_eq} becomes 
\begin{equation}
\ddot{\Delta}_\kappa + \Theta \Delta_\kappa = 0,
\label{pert_Eq_ESU}
\end{equation}
with
\begin{equation}
\Theta = \frac{(4 \alpha +8 C+8) \left(6 c_s^2+3 \omega ^2-8 \omega -1\right)}{16 (\omega +1)}+\frac{1}{4} \kappa ^2 c_s^2+3 \left(\omega -c_s^2\right),
\label{theta_1}
\end{equation}
where $\kappa$ is the comoving index which appears through the substitution $\mathrm{D}^2 \to - \frac{\kappa^2}{a_s^2}$. Note that the comoving index evolves in discrete form as $ \kappa^2 = \gamma(\gamma+2)$ for closed Universe case, while it evolves continuously like $\kappa^2 = \gamma^2 +1$, where $\gamma$ denotes the comoving wave number \cite{Harrison1967Oct}. Also, as shown by Barrow et al \cite{Barrow2003May}, $\gamma$ (represented by $n$ in their paper. We have avoided using $n$ as it already denotes the topological defect dimensionality in this present context) takes on values $\ge 2$ for physical inhomogeneous modes. Thus, the simplest case of $\gamma = 2$ gives $\kappa = 2.45$ and $2.23$ for closed and open Universe respectively. These values will be used to study the inhomogeneous density perturbations.

Similarly, for the open Universe case $k = -1$ case we have
\begin{equation}
\Theta = \frac{(4 \alpha +8 C-8) \left(6 c_s^2+3 \omega ^2-8 \omega -1\right)}{16 (\omega +1)}+\frac{1}{4} \kappa ^2 c_s^2+3 \left(\omega -c_s^2\right).
\label{Theta_k=-1}
\end{equation}
Now, it is evident that Eq. \eqref{pert_Eq_ESU} is a second-order differential equation that gives finite sinusoidal solutions provided $\Theta > 0$. The existence intervals in this case are complicated inequalities as there are five parameters of dependency, and so we have avoided expressing it explicitly. However, we can solve Eq. \eqref{pert_Eq_ESU} for $\Delta$ numerically for different combinations of the parameters.
\begin{figure*}
\centerline{\includegraphics[scale=0.4]{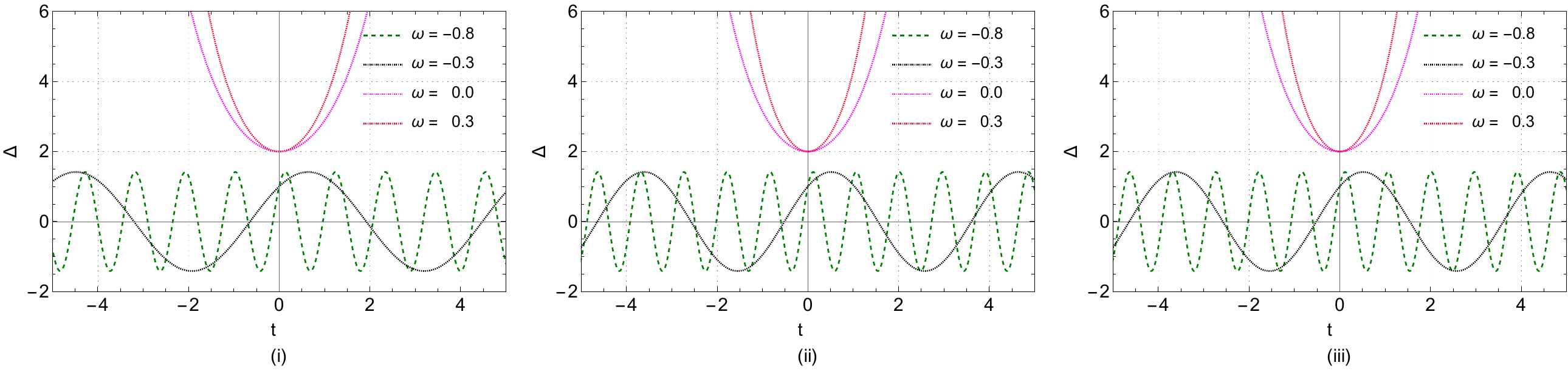}}
\caption{The evolution of density perturbation is shown for different values of $\omega$ for $k  = 1$ case. Here we have assumed$(\alpha, C, c_s) = (0.5, 0.6, 0.2)$ and (i) , (ii) and  (iii) represents the cases for monopoles, cosmic strings and domain walls respectively. Here the comoving index is $\kappa = 2.45$, for closed Universe.}
\label{dens_pert_fig}
\end{figure*}
\begin{figure*}
\centerline{\includegraphics[scale=0.4]{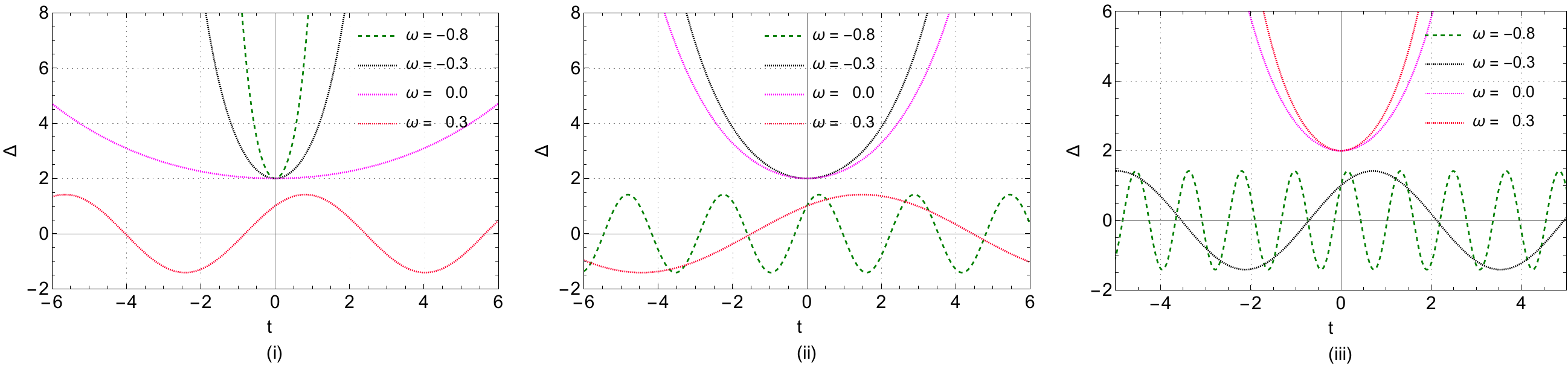}}
\caption{The evolution of density perturbation is shown for different values of $\omega$ for $k  = -1$ case. Here, too we have assumed$(\alpha, C, c_s) = (0.5, 0.6, 0.2)$ and  (i) , (ii) and  (iii) represents the cases for monopoles, cosmic strings and domain walls respectively. Here the comoving index is $\kappa = 2.23$, for open Universe.}
\label{dens_pert_fig_k-1}
\end{figure*}

\begin{figure*}
	\centerline{\includegraphics[scale=0.4]{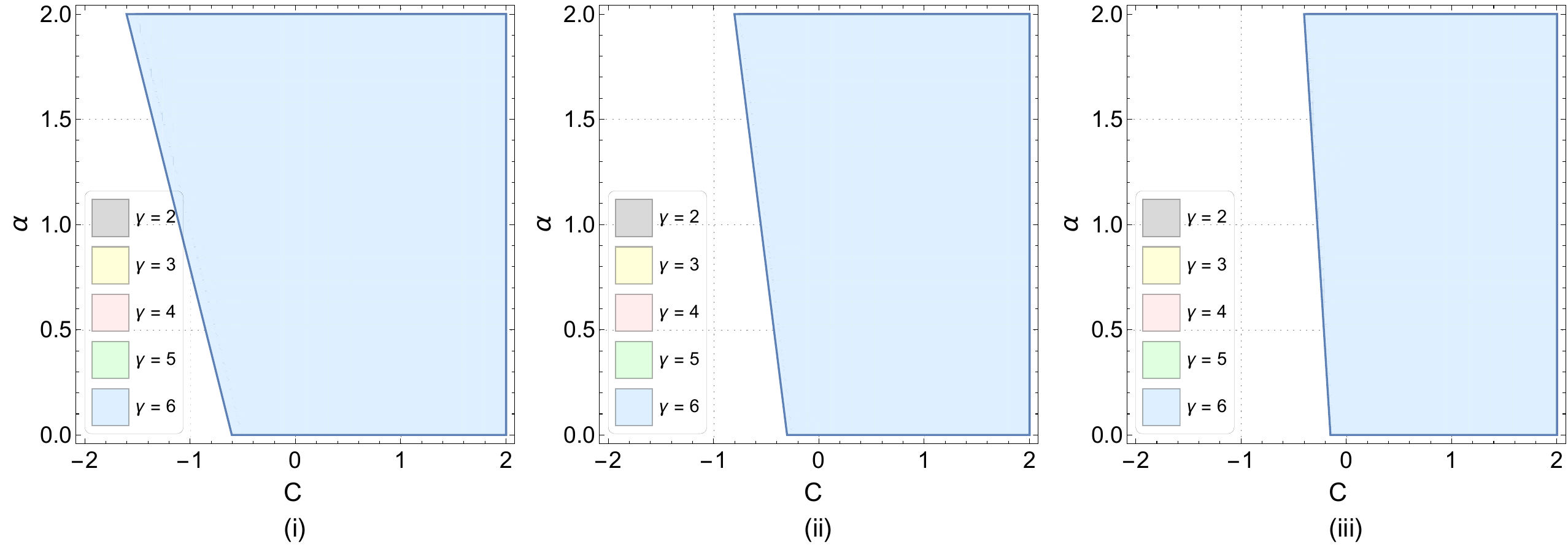}}
	\caption{The region plot shows the parameter space for allowed values required for stability of the ESU under inhomogeneous density perturbation for closed Universe. Here we have assumed a fixed value of sound speed $c_s^2 = 0.2$ and $\omega = -0.3$. }
	\label{dens_pert_fig_reg_k1}
\end{figure*}
\begin{figure*}
	\centerline{\includegraphics[scale=0.4]{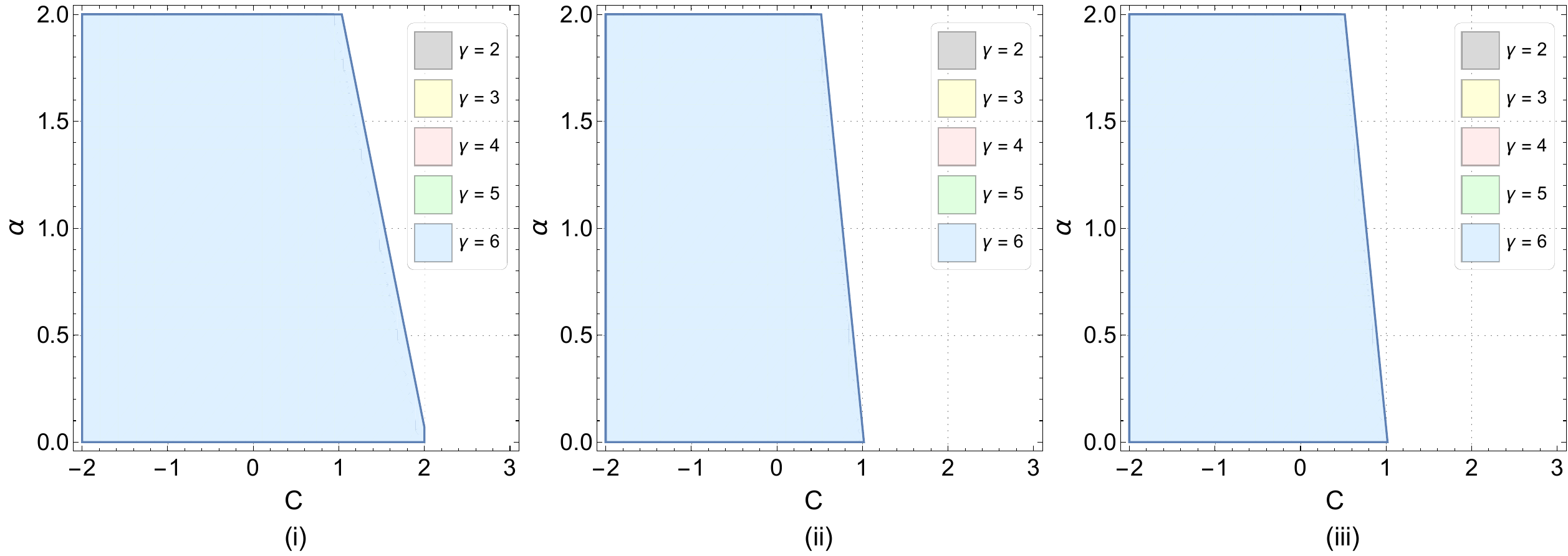}}
	\caption{The region plot shows the parameter space for allowed values required for stability of the ESU under inhomogeneous density perturbation for open Universe. Here too for stable ESU, we have assumed a fixed value of sound speed $c_s^2 = 0.2$ and $\omega = 0.3$ for monopoles and cosmic string cases (shown in (i) and (ii) respectively). However, for domain walls the value of $\omega$ is chosen to be  $-0.3$ for stable ESU.}
	\label{dens_pert_fig_reg_k-1} 
\end{figure*}

Now for the case of closed Universe, as seen from Fig. \ref{dens_pert_fig}, negative values of the EoS parameter $\omega$ ($\omega = -0.8, -0.3$) lead to stable and finite oscillatory modes in the presence of monopoles but the density perturbation blows up for positive values of $\omega$. This signify exotic fluid is a requirement for stable ESU, subject to inhomogeneous density perturbations. Similar observation is made in the presence of cosmic strings and domain walls. However, contrasting behaviour is seen in the case of the open Universe where a perfect fluid with $\omega > 0$, in the presence of monopoles lead to a stable ESU under inhomogeneous density perturbations. An interesting behaviour can be observed in the presence of cosmic strings wherein for $\omega > 0$, stable and finite oscillations of density perturbation exist, implying a stable ESU. Whereas, for $\omega = 0$ and $-0.3$, the perturbation diverges indicating an unstable ESU. But, for $-0.3 < \omega < -1$, the perturbations remain finite and hence a stable ESU. Finally, in the presence of domain walls in an open Universe, the behaviour is closely similar to that of the monopoles in a closed Universe. In essence, it is noticed that there is a very interesting interplay between the spatial curvature $k$ and the dimension of topological defect $n$ in determining the stability of the ESU under inhomogeneous density perturbations. 

Following \cite{Bohmer2015Dec}, we present Figs. \ref{dens_pert_fig_reg_k1} and \ref{dens_pert_fig_reg_k-1} that show the parameter space for the stable ESU under inhomogeneous density perturbation for $k = 1$ and $k = -1$ respectively. It is observed that the allowed values of the parameters $C$ and $\alpha$ overlap for the chosen values of $\gamma = 2, 3, 4, 5$ and $6$ indicating independence from the value of $\gamma$. So it can be concluded that the ESU is stable for all possible values of the comoving wave number $\gamma$.

\section{Stability under Vector and Tensor Perturbations}
\label{vec_tens_pert}
\subsection{Vector Perturbation}
The vector perturbations of a perfect fluid in a FLRW background are determined by the comoving dimensionless vorticity, which is defined as $\varpi_a = a \varpi$. The propagation equation for the perturbation modes is as follows. \cite{Barrow2003May}
\begin{equation}
\dot{\varpi}_k + H (1 - 3c_s^2)\varpi_k = 0,
\label{vect_pert_1}
\end{equation}
with $H$ as the Hubble parameter. In an ESU, as $H = 0$ Eq. \eqref{vect_pert_1} becomes
\begin{equation}
\dot{\varpi}_k = 0,
\label{vect_pert_2}
\end{equation}
The initial vector perturbations are clearly frozen, which is a basic result from Eq. \eqref{vect_pert_2}.  As a result, the ESU is neutrally stable against vector perturbation over all possible scales and EoS values.
\subsection{Tensor Perturbations}
The comoving dimensionless transverse-traceless shear is defined by $\Sigma_{ab} = a \sigma_{ab}$, corresponding to tensor perturbations, or gravitational wave perturbations associated with a perfect fluid having EoS $p=\omega \rho$, with modes fulfilling the following evolution equation (in the absence of $\Lambda$) \cite{Barrow2003May}:
\begin{equation}
\ddot{\Sigma}_k + 3H \dot{\Sigma}_k + \left[\frac{\kappa^2}{a^2} + \frac{2k}{a^2} - \frac{(1+3\omega)\rho}{3}\right]\Sigma_k =0,
\label{tens_pert}
\end{equation}
Like the previous analysis, we set $a = a_s = 2$ and $\rho = \rho_s$ and then using Eq. \eqref{rhos_gen}, Eq. \eqref{tens_pert} becomes
\begin{equation}
\ddot{\Sigma}_k + \Phi \Sigma_k = 0,
\label{tens_pert_2}
\end{equation}
where 
\begin{equation}
\Phi = \frac{3}{4}-\frac{(3 \omega +1) \left(\alpha + 2^{n+1} C+2\right)}{48 \pi  (\omega +1)},
\label{tens_per_k1}
\end{equation}
and 
\begin{equation}
\Phi = \frac{3}{4}-\frac{(3 \omega +1) \left(\alpha + 2^{n+1} C-2\right)}{48 \pi  (\omega +1)},
\label{tens_per_k-1}
\end{equation}
for closed and open Universe respectively.
Eq. \eqref{tens_pert_2} determines the neutral stability of tensor perturbation. Stable and unstable oscillation modes are specified by the conditions $\Phi > 0$ and $\Phi < 0$ respectively. 

\begin{figure*}
\centerline
\centerline{\includegraphics[scale=0.4]{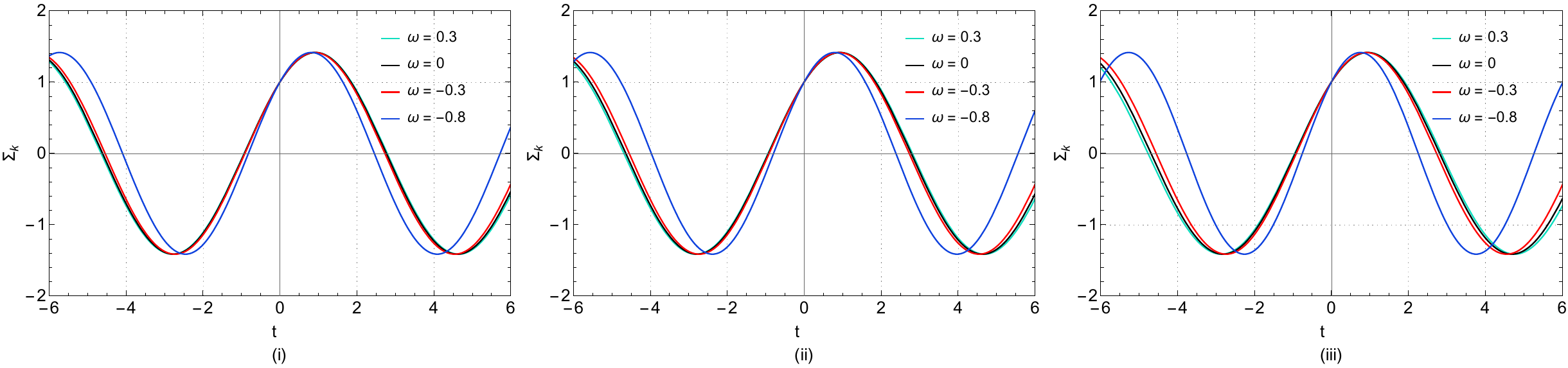}}
\caption{The evolution of tensor perturbation is shown for different values of $\omega$. Here we have assumed $(\alpha, C) = (0.5, 0.6)$ for $k = 1$. Also (i), (ii), and (iii) represent monopole, cosmic string and domain wall cases.}
\label{tens_pert_figk1}
\end{figure*}
\begin{figure*}
\centerline
\centerline{\includegraphics[scale=0.4]{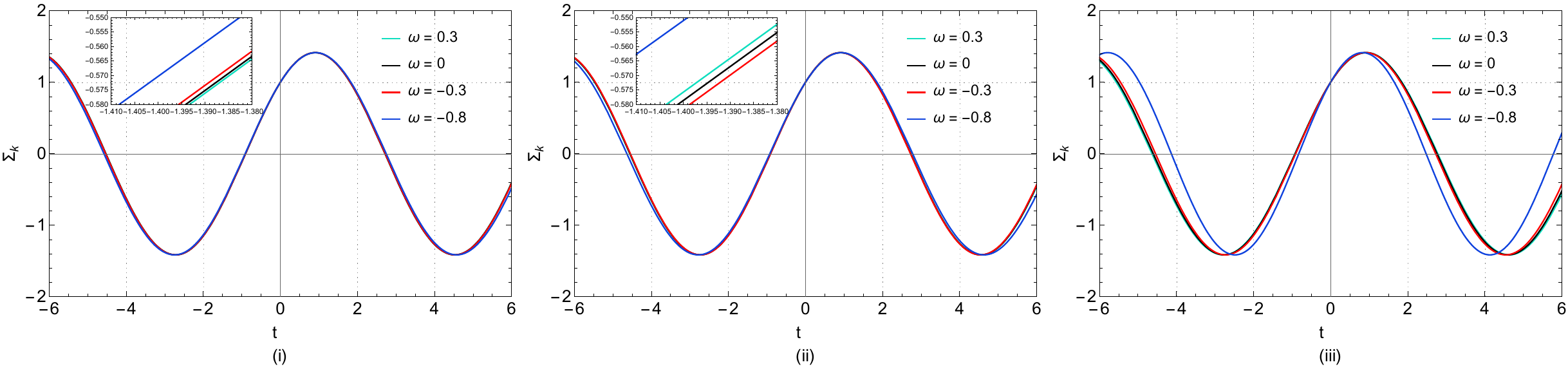}}
\caption{The evolution of tensor perturbation is shown for different values of $\omega$. Here we have assumed $(\alpha, C) = (0.5, 0.6)$ for $k = -1$. Also (i), (ii), and (iii) represent monopole, cosmic string and domain wall cases.}
\label{tens_pert_figk-1}
\end{figure*}

The evolutionary behaviour of the tensor perturbations is shown in Figs. \ref{tens_pert_figk1} and \ref{tens_pert_figk-1}. It is clearly observed that the tensor perturbations for both closed and open Universes remain finite for both positive and negative $\omega$ values regardless of the dimension of topological defect. Therefore, the ESU exhibits stable behavior subject to tensor perturbations in the presence of monopoles, cosmic strings, and domain walls, regardless of whether the perfect fluid obeys or violates the SEC.

The existence intervals in terms of tensor perturbation for closed Universe is explained as follows. Let us for define the following for ease of understanding:
\[
\omega_{\mathrm{crit}(1)} =
\begin{cases}
\displaystyle\frac{-2-2C+36\pi-\alpha}{6+6C-36\pi+3\alpha}, & \text{(closed)} \\[1mm]
\displaystyle\frac{2-2C+36\pi-\alpha}{-6+6C-36\pi+3\alpha}, & \text{(open)}
\end{cases}
\]
\[
\omega_{\mathrm{crit}(2)} =
\begin{cases}
\displaystyle\frac{-2-8C+36\pi-\alpha}{6+24C-36\pi+3\alpha}, & \text{(closed)} \\[1mm]
\displaystyle\frac{2-8C+36\pi-\alpha}{-6+24C-36\pi+3\alpha}, & \text{(open)}
\end{cases}
\]

\bigskip
\noindent
\textbf{In a Closed Universe $(k = 1)$}, we have the following existence regions for $\omega$ for different kinds of topological defects:
\medskip
\noindent
\textbf{For Monopoles ($n=0$):}
\begin{itemize}
    \item Provided \(0<\alpha<-2+12\pi\): With
    \(
    0<C\le \frac{1}{2}(-2+12\pi-\alpha),
    \) one requires 
    \(
    \omega > -1.
    \)
    \item When \(\alpha\ge -2+12\pi\): With \(C>0\), one requires
    \(
    -1<\omega<\omega_{\mathrm{crit}(1)}.
    \)
\end{itemize}

\noindent
\textbf{For Cosmic Strings ($n=1$):}
\begin{itemize}
    \item When \(0<\alpha<-2+12\pi\): If 
    \(
    0<C\le \frac{-2+12\pi-\alpha}{2},
    \)
    then one must have 
    \(
    \omega > -1.
    \)
    \item When \(\alpha\ge -2+12\pi\): With \(C>0\), we have
    \(
    -1<\omega<\omega_{\mathrm{crit}(1)}.
    \)
\end{itemize}

\noindent
\textbf{Domain Walls ($n=2$):}
\begin{itemize}
    \item Provided \(0<\alpha<-2+12\pi\): If 
    \(
    0<C\le \frac{-2+12\pi-\alpha}{8},
    \)
    then we require 
    \(
    \omega > -1.
    \)
    \item When \(\alpha\ge -2+12\pi\): With \(C>0\), we one requires
    \(
    -1<\omega<\omega_{\mathrm{crit}(2)}.
    \)
\end{itemize}

\bigskip
\noindent
\textbf{In an Open Universe $(k = -1)$}, similar to the previous case, we can have the existence regions for $\omega$ for all kinds of topological defects as follows:
\medskip
\noindent

\textbf{For Monopoles ($n=0$) \& Cosmic Strings ($n=1$):}

\textbf{Case I:} Provided \(0<\alpha<2\)
\begin{itemize}
    \item For \(0<C<\frac{2-\alpha}{2}\): the constraint on $\omega$ is
    \(
    \omega < -1 \quad \text{or} \quad \omega > \omega_{\mathrm{crit}(1)}.
    \)
    \item For \(C=\frac{2-\alpha}{2}\): No constraint on \(\omega\).
    \item For \(\frac{2-\alpha}{2}<C<\frac{1}{2}(2+12\pi-\alpha)\): we must have 
    \(
    \omega < \omega_{\mathrm{crit}(1)} \quad \text{or} \quad \omega > -1.
    \)
\end{itemize}

\textbf{Case II:} Provided \(2\le\alpha<2+12\pi\)
\begin{itemize}
    \item For \(0<C<\frac{1}{2}(2+12\pi-\alpha)\): The constraint on $\omega$ is
    \(
    \omega < \omega_{\mathrm{crit}(1)} \quad \text{or} \quad \omega > -1.
    \)
\end{itemize}

\textbf{Case III:} Provided \(\alpha\ge2+12\pi\), with
\(
C>0 \text{ one must have } -1<\omega<\omega_{\mathrm{crit}(1)}.
\)

\medskip
\noindent
\textbf{Domain Walls ($n=2$):}

\textbf{Case I:} Provided \(0<\alpha<2\)
\begin{itemize}
    \item For \(0<C<\frac{2-\alpha}{8}\): The constraint on $\omega$ is
    \(
    \omega < -1 \quad \text{or} \quad \omega > \omega_{\mathrm{crit}(2)}.
    \)
    \item For \(C=\frac{2-\alpha}{8}\): No constraint on \(\omega\).
    \item For \(\frac{2-\alpha}{8}<C<\frac{2+12\pi-\alpha}{8}\): One must have
    \(
    \omega < \omega_{\mathrm{crit}(2)} \quad \text{or} \quad \omega > -1.
    \)
\end{itemize}

\textbf{Case II:} Provided \(2\le\alpha<2+12\pi\)
\begin{itemize}
    \item For \(0<C<\frac{2+12\pi-\alpha}{8}\): The constraint on $\omega$ is 
    \(
    \omega < \omega_{\mathrm{crit}(2)} \quad \text{or} \quad \omega > -1.
    \)
\end{itemize}

\textbf{Case III:} Provided \(\alpha\ge2+12\pi\), with
\(
C>0 \text{ we should have the constraint on } \omega \text{ as }  -1<\omega<\omega_{\mathrm{crit}(2)}.
\)

\section{Conclusion}
\label{conc}
In order to understand the essential conditions for such a potential early Universe scenario—namely, \emph{stability of the ESU} and \emph{graceful exit mechanism} into the conventional inflationary Universe—we have examined the emergent Universe scenario in this study.  The design of zero-point length cosmology allows us to study the ESU that occurs before to inflation by avoiding the initial singularity in the very early Universe.  By allowing for the non-zero initial size of the ESU, the finite length setup of a zero-point length universe may inherently satisfy the requirements of an EU scenario. The EU must exit the ESU and enter into the inflationary Universe successfully (graceful exit mechanism). The entry into inflation typically requires violation of SEC. To achieve this, it is usually expected that the total matter content of the Universe include exotic fluids that violate the SEC. As they do, they might play an important role in the transition from the ESU to an inflationary Universe. Some models achieve this SEC violation required for the phase transition from modified gravity as well. However, in this work we follow an alternative route: we investigate whether early Universe relics like topological defects might act as the exotic fluid and being reponsible for the ESU-to-inflationary phase transition.  As topological defects are interesting relics in the early Universe, that possess energy density, they can be incorporated as a part of the total energy density. 

In this work, we developed the modified Friedmann equations of a cosmological model based on a zero-point length framework that permits the existence of $n$-dimensional topological defects in the Universe's total matter content.  We examined the dynamical behavior of the system's eigen-valued squared by reducing the Friedmann equations to a set of autonomous differential equations. We carried out the analysis through taking into account two distinct scenarios: a closed and an open spatial geometry with 0, 1, and 2-dimensional topological defects, such as monopoles, cosmic strings, and domain wall defects, respectively. In this setting, we have observed a smooth graceful exit process for all the considered topological defects for different combination of the model parameters $\alpha$ and $C$. Notably, the dynamics of this exit process showed no drastic deviation  with the variation in the values of model parameter $C$. In contrast, $\lambda^2$ is found to be quite sensitive to the values of $\alpha$. This indicates that the closed Universe setting supports a successful graceful exit with no strict restriction on the values of $C$ and a noticeable constraint on $\alpha$. Thus higher values of $\alpha$ rules out the possibility of successful graceful exit in the closed Universe setting. However, in the open Universe scenario, we notice a similar occurrence of a successful graceful exit from the stable ESU to the inflationary phase, but with no noticeable sensitivity to the model parameter $\alpha$ and $C$.

Next, the stability of the ESU is carried out with respect to scalar homogeneous perturbation to the scale factor and energy density. The perturbation evolved as stable sinusoidal oscillations eternally in the ES phase. However, it appears the break from the ESU to the inflationary phase is sensitive to the values of the EoS of the perfect fluid of the model. To exit into the standard cosmology, the stability of the ESU should break and enter the inflationary phase. It turns out that a SEC violating fluid is necessary to realize the graceful exit subject to scalar homogeneous perturbation for monopoles and cosmic strings. However, in the case of domain walls, there is no apparent transition to inflation and the ESU remains constant.  The ESU remains in a stable phase if the perfect fluid violates the SEC, which is the opposite of what happens in the closed Universe. In the case of an open ESU, a successful graceful exit is only seen when SEC is obeyed, i.e., $\omega > 0$. Also, from the perspective of a closed ESU, when we look into the context of the late-time accelerated expansion Universe of standard cosmology, in which a transition from the matter era to the dark energy era occurs as the EoS switches from $\omega \ge 0$ to $\omega < 0$; our model shows analogous behaviour when the transition from the ESU to the inflationary phase proceeds through a switching of the EoS.

In terms of inhomogeneous density perturbation, the stability of the ESU, in the case of a closed Universe, for fluids satisfying the SEC, for all cases of topological defects, the density perturbations diverge making the ESU unstable. However, a SEC violating fluid can stabilize the ESU. In an open Universe, a contrasting behaviour is found where the perfect fluid with $\omega > 0$, in the presence of monopoles lead to a stable ESU. In the presence of cosmic strings $\omega > 0$ and $\omega < -1/3$ leads to a a stable ESU. But for the case of domain walls in an open Universe, the evolution of perturbation is similar to that of the monopoles in the case of a closed Universe. Thus, higher spatial curvature and lower dimensional topological defect behave similar to lower spatial curvature and higher dimensional topological defect in terms of inhomogeneous density perturbation. Vector perturbations are seen to remain frozen in the ESU, making it stable across all possible scales. Tensor perturbations, on the other hand, indicate that the ESU remains stable regardless of whether the fluid obeys or violates the SEC in both closed and open Universes. 

In this present work, we have addressed the stability and graceful exit in an emergent scenario and showed how variations in the EoS parameter and model parameters affect the process. However, we have not found an exact reason as to why the switching of the EoS parameter initiates the onset of inflation and is beyond the scope of the current analysis. This remains an open question. It would be interesting to address this question as  a future perspective.

\bibliography{bibliography.bib}
\end{document}